\pdfoutput=1
\documentclass[12pt]{article}
\usepackage[utf8]{inputenc}
\usepackage[british]{babel}
\usepackage{cmap}
\usepackage{lmodern}

\usepackage{amssymb, amsmath, amsthm}
\usepackage[a4paper,top=25mm,bottom=25mm,left=25mm,right=25mm]{geometry}
\usepackage{ragged2e}

\usepackage{authblk} 
\usepackage{pifont}
\usepackage{graphicx}
\usepackage[dvipsnames,svgnames,table]{xcolor}
\usepackage[figuresright]{rotating}
\usepackage{xtab} 
\usepackage{longtable} 
\usepackage{multirow}
\usepackage{footnote}
\usepackage[stable]{footmisc}
\usepackage{chngpage} 
\usepackage{pdflscape} 
\usepackage[nottoc,notlot,notlof]{tocbibind} 

\usepackage{pgfplots}
\pgfplotsset{compat=1.18}
\pgfplotsset{every tick label/.append style={font=\footnotesize}}
\usepackage{setspace}

\makesavenoteenv{tabular}
\usepackage{tabularx}
\usepackage{booktabs}
\usepackage{threeparttable} 
\usepackage[referable]{threeparttablex} 
\newcolumntype{R}{>{\raggedleft\arraybackslash}X}
\newcolumntype{L}{>{\raggedright\arraybackslash}X}
\newcolumntype{C}{>{\centering\arraybackslash}X}
\newcolumntype{A}{>{\columncolor{gray!25}}C}
\newcolumntype{a}{>{\columncolor{gray!25}}c}

\newlength{\tablen}

\usepackage{dcolumn} 
\newcolumntype{.}{D{.}{.}{-1}}

\usepackage{tikz}
\usetikzlibrary{arrows, calc, matrix, patterns, positioning, shapes, trees}
\usepackage[semicolon]{natbib}
\usepackage[hyphens]{url}
\usepackage{hyperref} 
\hypersetup{
  colorlinks   = true,    
  urlcolor     = blue,    
  linkcolor    = blue,    
  citecolor    = ForestGreen      
}
\usepackage{microtype}
\usepackage[singlelinecheck=false,justification=centering]{caption} 

\usepackage[labelformat=simple]{subcaption}

\DeclareCaptionLabelFormat{parenthesis}{(#2)}
\makeatletter
\renewcommand\p@subfigure{\arabic{figure}.}
\makeatother

\DeclareCaptionLabelFormat{parenthesis}{(#2)}
\makeatletter
\renewcommand\p@subtable{\arabic{table}.}
\makeatother

\usepackage{enumitem}
\setlist[itemize]{leftmargin=2.5\parindent}
\setlist[enumerate]{leftmargin=2.5\parindent}

\def\addlegendimage{\csname pgfplots@addlegendimage\endcsname}

\theoremstyle{plain}

\theoremstyle{definition}

\newtheorem{example}{Example}

\theoremstyle{remark}

\def\keywords{\vspace{.5em} 
{\noindent \textit{Keywords}: }}

\def\JEL{\vspace{.5em} 
{\noindent \textbf{\emph{JEL} classification number}: }}

\def\AMS{\vspace{.5em} 
{\noindent \textbf{\emph{MSC} class}: }}

\title{The fairness of the group draw \\ for the FIFA World Cup}
\author{\href{https://sites.google.com/view/laszlocsato}{L\'aszl\'o Csat\'o}\thanks{~E-mail: \emph{laszlo.csato@sztaki.hun-ren.hu}} }
\affil{Institute for Computer Science and Control (SZTAKI) \\
Hungarian Research Network (HUN-REN) \\
Laboratory on Engineering and Management Intelligence \\
Research Group of Operations Research and Decision Systems}
\affil{Corvinus University of Budapest (BCE) \\
Institute of Operations and Decision Sciences \\
Department of Operations Research and Actuarial Sciences}
\affil{Budapest, Hungary}
\date{\today}

\def\Dedication{
{\noindent
``\emph{Any one who considers arithmetical methods of producing random digits is, of course, in a state of sin.}''\footnote{~Source: \citet[p.~36]{Neumann1951}.}
}

\flushright
(John von Neumann: \emph{Various techniques used in connection with random digits})

\vspace{0.5cm} 
\justify }

\begin{document}

\newgeometry{top=25mm,bottom=25mm,left=25mm,right=25mm}

\maketitle
\thispagestyle{empty}
\Dedication

\begin{abstract}
\noindent
Several sports tournaments contain a round-robin group stage where the teams are assigned to groups subject to some constraints. Hence, the organisers usually use a computer-assisted random draw to avoid any dead end, a situation when the teams still to be drawn cannot be assigned to the remaining empty slots. This procedure is known to be unfair: the feasible allocations are not equally likely, that is, the draw does not have a uniform distribution. We quantify the implied unfairness of the 2018 FIFA World Cup draw and evaluate its effect on the probability of qualification for the knockout stage for each national team. The official draw order of Pot 1, Pot 2, Pot 3, Pot 4 turns out to be a significantly better option than the 23 other draw orders with respect to the unwanted distortions. Nonetheless, the non-uniform draw distorts the probability of qualification by more than one percentage point for two countries.
Our results call attention to the non-negligible role of draw order and make it possible for policymakers to decide whether using fairer draw mechanisms is justified.


\AMS{05A05, 68W40, 90-10, 91B14}

\JEL{C44, C63, Z20}
\end{abstract}

\clearpage
\restoregeometry

\section{Introduction} \label{Sec1}

The mechanism design literature usually focuses on theoretical requirements like efficiency, fairness, and incentive compatibility \citep{AbdulkadirougluSonmez2003, RothSonmezUnver2004, Csato2021a}. On the other hand, institutions---like governing bodies in major sports---often emphasise simplicity and transparency, which calls for a comprehensive review of how the procedures that exist in the real world perform with respect to the above properties.

Many sports tournaments involve a group stage where the teams are assigned to groups subject to some constraints \citep{Kobierecki2022, LalienaLopez2019}.
These constraints are imposed ``\emph{to issue a schedule that is fair for the participating teams, fulfils the expectations of commercial partners and ensures with a high degree of probability that the fixture can take place as scheduled}'' \citep{UEFA2020c}.
Examples include the FIBA Basketball World Cup \citep{FIBA2023}, the FIFA World Cup \citep{FIFA2022a}, the European Qualifiers for the FIFA World Cup \citep{UEFA2020c}, the UEFA Euro qualifying \citep{UEFA2022e}, and the UEFA Nations League \citep{UEFA2024a}.
Draw restrictions are also used in club-level tournaments such as the UEFA Champions League and the UEFA Europa League \citep{Csato2022d}.

Until 2014, the FIFA World Cup draw divided the teams into pots mainly according to their geographic area, which caused serious unfairness in certain years such as in 1990 \citep{Jones1990}, 2006 \citep{RathgeberRathgeber2007}, and 2014 \citep{Guyon2015a}. In particular, the top teams had different chances of being placed in a group with
difficult opponents. Consequently, \citet{Guyon2014a} has recommended three options to create balanced and geographically diverse groups, which has inspired FIFA to change the draw procedure used in the 2018 and 2022 FIFA World Cups \citep{Guyon2018d, Csato2023d}.

However, this mechanism, based on a computer-assisted random draw to avoid any dead end (a situation when the teams still to be drawn cannot be assigned to the remaining empty slots), is not uniformly distributed: the valid assignments are not equally likely to occur \citep{RobertsRosenthal2024}.
Hence, \citet{RobertsRosenthal2024} propose two unbiased mechanisms that use balls and bowls, making them suitable for a televised draw. On the other hand, they require random simulations, which might threaten transparency.

Therefore, the organiser faces a dilemma: retain the existing but biased method or switch to a correct but less transparent mechanism. In order to understand this trade-off and choose the better option, it is inevitable to explore the extent of the bias, as well as its potential sporting effects. While these issues have been recently analysed in the UEFA Champions League \citep{BoczonWilson2023}, they have not been addressed in the case of the FIFA World Cup, the most prominent football tournament around the world.

The current work aims to fill this research gap by analysing the \emph{unique} mechanism used in practice by sports federations to draw groups (containing at least three teams) with constraints through the example of the 2018 FIFA World Cup. This case study has been chosen since the 2022 FIFA World Cup draw has contained an inherent bias due to the uncertainty in the set of teams qualified \citep{Csato2023d}, and the teams qualifying for the 2026 FIFA World Cup are naturally unknown at the time of writing.

We also call attention to the role of \emph{draw order}: the order of the pots in the group draw turns out to have a non-negligible sporting effect. This is a somewhat surprising finding because the fairness of the UEFA Champions League Round of 16 draw essentially does not depend on whether the group winners or the runners-up are drawn first \citep{BoczonWilson2023, KlossnerBecker2013}.
On the other hand, no clear recommendation is given for the optimal draw order in general due to the huge complexity of the calculations; according to \citet[Section~5.1]{Csato2024h}, computing the probabilities for six teams drawn from three pots is barely possible---and the 2018 FIFA World Cup contains 32 teams and eight pots. Therefore, several more case studies would be needed to find a general pattern if it exists.

Our topic is hugely relevant to governing bodies in sports.
In an ideal environment, the rules governing a sport prevent any player or team from gaining an unfair advantage \citep{Csato2021a, DevriesereCsatoGoossens2024}. But the sequential draw method of FIFA may threaten fairness, while the alternative solutions remain more complex and less transparent \citep{RobertsRosenthal2024}. Therefore, policymakers could maintain the balance between fairness and integrity only by investigating different fairness indicators \citep{ScellesFrancoisValenti2024}.

The paper starts with presenting a concise review of related literature, followed by describing the theoretical background of the FIFA World Cup group draw.
The numerical results for the 2018 FIFA World Cup draw are discussed in two parts. First, the departure of the draw procedure from a uniformly distributed random choice among all feasible allocations is quantified for all the 24 possible draw orders of the four pots. This is important because the modification of the draw order (relabelling of the pots) does not require any reform in the existing principles of the draw.
Second, the bias of the draw procedure is evaluated with respect to the probability of qualification for the knockout stage. That is essential since the \emph{ultimate} price to pay for the sake of public interest and transparency is the distortion of the final outcome: if the effects on the chances of the teams remain marginal and insignificant, then there is no need to choose a more complex and less transparent draw procedure.

\section{Related literature} \label{Sec2}

Several scientific works analyse the FIFA World Cup draw. Before the 2018 edition, the host nation and the strongest teams were assigned to different groups, while the remaining teams were drawn randomly with maximising geographic separation: countries from the same continent (except for Europe) could not have played in the same group and at most two European teams could have been in any group.

In the case of the 1990 FIFA World Cup, \citet{Jones1990} shows that the draw was not mathematically fair. For example, West Germany would have been up against a South American team with a probability of 4/5 instead of 1/2---as it should have been---due to the incorrect consideration of the constraints.
Similarly, the host Germany was likely to play in a difficult group in 2006, but other seeded teams, such as Italy, were not \citep{RathgeberRathgeber2007}.

\citet{Guyon2015a} identifies severe shortcomings of the mechanism used for the 2014 FIFA World Cup draw such as imbalance (the eight groups are at different competitive levels), unfairness (certain teams have a greater chance of ending up in a tough group), and non-uniform distribution (the feasible allocations are not equally likely). Indeed, there has been a substantial competitive imbalance between the historical FIFA World Cup groups \citep{LaprePalazzolo2023}.

\citet{Guyon2014a} presents alternative proposals to retain the practicalities of the FIFA World Cup draw but improve its outcome. One of them can be compared to the flawed FIFA rule at \url{https://www.nytimes.com/interactive/2014/06/03/upshot/world-cup-draw-simulation.html}.

\citet{LalienaLopez2019} develop two uniformly distributed designs for group draw with geographical restrictions that produce groups having similar or equal competitive levels.
\citet{CeaDuranGuajardoSureSiebertZamorano2020} analyse the deficiencies of the 2014 FIFA World Cup draw and provide a mixed integer linear programming model to create the groups. The suggested method takes draw restrictions into account and aims to balance ``quality'' across the groups.

\citet{RobertsRosenthal2024} consider the challenge of finding a group draw mechanism that follows the uniform distribution over all valid assignments but is also entertaining, practical, and transparent. The authors suggest two procedures for achieving this aim by using balls and bowls in a way, which is suitable for a nice television show---but, in contrast to the proposals of \citet{Guyon2014a}, they use computer draws at some stage which may threaten transparency. Both algorithms can be tried interactively at \url{http://probability.ca/fdraw/}.

Other studies deal with the UEFA Champions League where a constrained draw mechanism has been used in the Round of 16 between the 2003/04 and 2023/24 seasons. However, this method is distinct from the FIFA World Cup draw procedure \citep{Csato2024h}.
\citet{WallaceHaigh2013} verify that the possible assignments are not equally likely and highlight the connection of the UEFA draw mechanism to Hall's marriage theorem, see also \citet[Section~3.6]{Haigh2019}.
\citet{Kiesl2013} computes the bias in the 2012/13 season and outlines some fair---but uninteresting to watch---methods.
According to \citet{KlossnerBecker2013}, the draw system inherently implies different probabilities for certain assignments, which are translated into more than ten thousand Euros in expected revenue due to the substantial amount of prize money.
Finally, \citet{BoczonWilson2023} reveal how the UEFA draw procedure affects expected assignments and address the normative question of whether a fairer randomisation mechanism exists. The current design is verified to come quantitatively close to a constrained best in fairness terms.

\section{Theoretical background} \label{Sec3}

A \emph{permutation} of a set is a rearrangement of its elements. In the FIFA World Cup draw, the initial permutation of the teams is provided by a random draw. In an unrestricted group draw, the teams can be assigned to the groups in this permutation. However, in the presence of draw conditions, it is not obvious to find the permutation of the teams that corresponds to the feasible allocation implied by the draw procedure of FIFA.

This mechanism is defined as follows: ``\emph{when a draw condition applies or is anticipated to apply, the team drawn is allocated to the first available group in alphabetical order}'' \citep{UEFA2020c}. In other words, the team drawn is assigned to the first empty slot except if all permutations of the remaining teams violate at least one draw condition.

\input{Figure1_FIFA_permutation}

\begin{example} \label{Examp1}
Assume that there are $k=4$ groups A--D and $n=4$ teams $T1$--$T4$ drawn sequentially from a pot. The sequence of permutations to be checked according to the draw mechanism is shown in Figure~\ref{Fig1}. Team $T1$ is assigned to group A in the first six permutations because it can be placed in another group only if either group A is unavailable for team $T1$ or teams $T2$--$T4$ cannot be assigned to groups B--D.

Let us consider two illustrative cases, where the restrictions are implied by the assignment of teams drawn from the previous pot(s):
\begin{itemize}
\item
If team $T1$ cannot be placed in group A and team $T3$ cannot be placed in group C, then the first six permutations are unacceptable due to the first constraint, and permutation 7 is skipped because of the second condition.
The outcome of the draw procedure is permutation 8 ($T2$, $T1$, $T4$, $T3$).

The draw procedure of FIFA works as follows:
\begin{itemize}[label=$\diamond$]
\item
Team $T1$ cannot be assigned to the first empty slot in group A, hence, it is placed in the first available group, which is group B.
\item
Team $T2$ is assigned to the first empty slot in group A.
\item
Team $T3$ cannot be assigned to the first empty slot in group C, hence, it is placed in the first available group, which is group D.
\item
Team $T4$ is assigned to the first (and last) empty slot in group C.
\end{itemize}

\item
If teams $T2$--$T4$ cannot be placed in group C and team $T2$ cannot be placed in group A, then the first 12 permutations are unacceptable due to the first constraint, and the next two are skipped because of the second condition.
The outcome of the draw procedure is permutation 15 ($T3$, $T2$, $T1$, $T4$).

The draw procedure of FIFA works as follows:
\begin{itemize}[label=$\diamond$]
\item
No constraint prohibits directly the assignment of team $T1$ to group A. However, if team $T1$ is not assigned to group C, then the three remaining teams $T2$--$T4$ cannot be assigned to groups B--D. A similar argument uncovers that team $T1$ should be assigned to group C.
\item
Team $T2$ cannot be assigned to the first empty slot in group A, hence, it is placed in the first available group, which is group B.
\item
Team $T3$ is assigned to the first empty slot in group A.
\item
Team $T4$ is assigned to the first (and last) empty slot in group D.
\end{itemize}
\end{itemize}
\end{example}

The official video of the 2018 FIFA World Cup group draw is available at \url{https://www.youtube.com/watch?v=jDkn83FwioA}.

Generating all permutations of a sequence of values is a famous problem in computer science \citep{Sedgewick1977}. The classic lexicographic algorithm goes back to \emph{Naraya{\d n}a Pa{\d n}{\d d}ita}, an Indian mathematician from the 14th century \citep{Knuth2005}. The sequence corresponding to the FIFA World Cup draw procedure is called \emph{representation via swaps} \citep{Arndt2010} and has been presented first in \citet{MyrvoldRuskey2001} according to our knowledge. In particular, Figure~\ref{Fig1} is analogous to \citet[Figure~10.1-E]{Arndt2010}.

\begin{figure}[t!]
\centering

\begin{tikzpicture}[scale=1,auto=center, transform shape, >=triangle 45]
\tikzstyle{every node}=[draw,align=center];
  \node (N1) at (-2,12) {START};
  \node[shape = ellipse] (N2) at (-2,9) {Can team $i$ of pot $j$ \\ be placed in group $k$?};
  \node (N3) at (-2,6) {Team $i$ of pot $j$ is \\ assigned to group $k$ \\ $k = 1$};
  \node[shape = ellipse] (N4) at (-2,3) {Is $i$ smaller than the \\ number of teams $n$?};
  \node (N5) at (-5,0) {$i = i + 1$ \\ START};
  \node (N6) at (-1.5,-0.78) {\textbf{\textcolor{ForestGreen}{The feasible}} \\ \textbf{\textcolor{ForestGreen}{allocation implied}} \\ \textbf{\textcolor{ForestGreen}{by the FIFA}} \\ \textbf{\textcolor{ForestGreen}{mechanism is}} \\ \textbf{\textcolor{ForestGreen}{found}}};
  \node[shape = ellipse] (N7) at (6,6) {Is $k$ smaller than \\ the number of groups \\ available for pot $j$?};
  \node (N8) at (3,3) {$k = k + 1$ \\ START};
  \node[shape = ellipse] (N9) at (7,3) {Is $i = 1$?};
  \node (N10) at (2.25,-0.28) {\textbf{\textcolor{BrickRed}{There exists}} \\ \textbf{\textcolor{BrickRed}{NO feasible}} \\ \textbf{\textcolor{BrickRed}{allocation}}};
  \node (N11) at (7,-0.78) {$k = \left[ \text{group of team } (i-1) \right] + 1$ \\ All teams $\ell \geq i$ are \\ removed from their groups \\ $i = i - 1$ \\ START};

\tikzstyle{every node}=[align=center];  
  \draw [->,line width=1pt] (N1) -- (N2)  node [midway, left] {};
  \draw [->,line width=1pt] (N2) -- (N3)  node [midway, left] {Yes};
  \draw [->,line width=1pt] (N3) -- (N4)  node [midway, left] {};
  \draw [->,line width=1pt] (N4) -- (N5)  node [midway, above left] {Yes};
  \draw [->,line width=1pt] (N4) -- (N6)  node [midway, above right] {No};
  \draw [->,line width=1pt] (N2) -- (N7)  node [midway, above right] {No};
  \draw [->,line width=1pt] (N7) -- (N8)  node [midway, above left] {Yes};
  \draw [->,line width=1pt] (N7) -- (N9)  node [midway, above right] {No};
  \draw [->,line width=1pt] (N9) -- (N10)  node [midway, above left] {Yes};
  \draw [->,line width=1pt] (N9) -- (N11)  node [midway, right] {No};
\end{tikzpicture}
\captionsetup{justification=centering}
\caption{A backtracking algorithm for restricted group draw that finds \\ the feasible allocation corresponding to a given order of the teams}
\label{Fig2}
\end{figure}
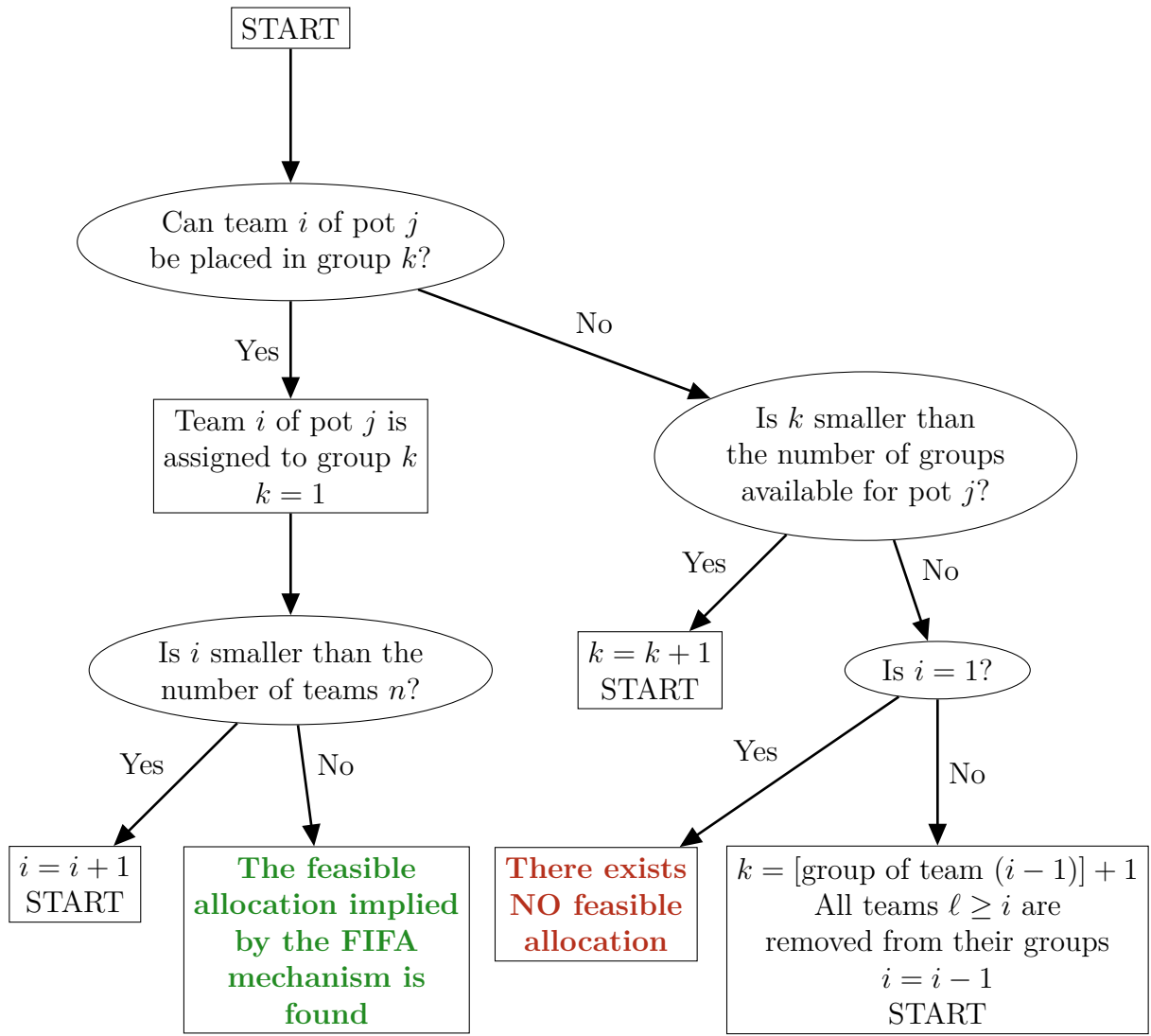


The description of the 2018 FIFA World Cup draw \citep{FIFA2017c} does not give an exact algorithm to obtain the implied feasible allocation of the teams into groups for a given random permutation of the teams.
The scheme of an appropriate computer program is presented in Figure~\ref{Fig2}. Pot $j$, from which team $i$ is drawn, is not a variable, however, the number of available groups can be different for each pot.
The algorithm is based on \emph{backtracking}: if the remaining teams cannot be assigned to the empty group slots in \emph{any} permutation such that all restrictions are satisfied, then the last team is placed in the next available group in alphabetical order. This process is repeated until the associated feasible allocation is obtained or the non-existence of a valid assignment is verified.

Backtracking can be familiar from the problem of scheduling round-robin tournaments, where an unlucky assignment of games to slots can result in a schedule that cannot be completed \citep{RosaWallis1982, Schaerf1999}.
Backtracking is also widely used to solve puzzles such as the eight queens puzzle, crosswords, or Sudoku. For draw procedures under draw constraints, backtracking has been suggested first in \citet{Guyon2014a}.

\section{The (un)fairness of the 2018 FIFA World Cup draw} \label{Sec4}

For the 2018 FIFA World Cup draw, pots were constructed based on the October 2017 FIFA World Ranking such that Pot $j$ contained the teams ranked between $8(j-1) + 1$ and $8j$. The only exception was the assignment of the host Russia to Pot 1 despite being the lowest-ranked among all participants.

The draw sequence started with Pot 1 and ended with Pot 4. Each pot was emptied before the next was drawn and some draw conditions were applied \citep{FIFA2017c}:
\begin{itemize}
\item
Russia was automatically placed in Group A.
\item
No group could have more than one team from any continental confederation except for UEFA (AFC, CAF, CONMEBOL, CONCACAF).
\item
Each group should have contained at least one but no more than two European teams.
\end{itemize}

\begin{table}[t!]
  \centering
  \caption{Seeding pots in the 2018 FIFA World Cup}
  \label{Table1}
    \rowcolors{1}{gray!20}{}
\begin{threeparttable}
    \begin{tabularx}{\textwidth}{Llc c Llc} \toprule \hiderowcolors
    Country & Confederation & Elo &       & Country & Confederation & Elo \\
    \midrule
    \multicolumn{3}{c}{\textbf{Pot 1}} &       & \multicolumn{3}{c}{\textbf{Pot 2}} \\ \bottomrule \showrowcolors
    1 Russia & UEFA  & 1678  &       & 9\textcolor{white}{0} Spain & UEFA  & 2044 \\
    2 Germany & UEFA  & 2077  &       & 10 Peru & CONMEBOL & 1916 \\
    3 Brazil & CONMEBOL & 2141  &       & 11 Switzerland & UEFA  & 1889 \\
    4 Portugal & UEFA  & 1969  &       & 12 England & UEFA  & 1948 \\
    5 Argentina & CONMEBOL & 1985  &       & 13 Colombia & CONMEBOL & 1927 \\
    6 Belgium & UEFA  & 1937  &       & 14 Mexico & CONCACAF & 1850 \\
    7 Poland & UEFA  & 1831  &       & 15 Uruguay & CONMEBOL & 1893 \\
    8 France & UEFA  & 1986  &       & 16 Croatia & UEFA  & 1853 \\ \toprule \hiderowcolors
    \multicolumn{3}{c}{\textbf{Pot 3}} &       & \multicolumn{3}{c}{\textbf{Pot 4
    }} \\ \bottomrule \showrowcolors
    17 Denmark & UEFA  & 1856  &       & 25 Serbia & UEFA  & 1777 \\
    18 Iceland & UEFA  & 1764  &       & 26 Nigeria & CAF   & 1684 \\
    19 Costa Rica & CONCACAF & 1743  &       & 27 Australia & AFC   & 1741 \\
    20 Sweden & UEFA  & 1795  &       & 28 Japan & AFC   & 1684 \\
    21 Tunisia & CAF   & 1655  &       & 29 Morocco & CAF   & 1733 \\
    22 Egypt & CAF   & 1643  &       & 30 Panama & CONCACAF & 1658 \\
    23 Senegal & CAF   & 1749  &       & 31 South Korea & AFC   & 1713 \\
    24 Iran & AFC   & 1790  &       & 32 Saudi Arabia & AFC   & 1586 \\
 \toprule
    \end{tabularx}
\begin{tablenotes} \footnotesize
\item
The number before each country indicates its rank among the FIFA World Cup participants according to the October 2017 FIFA World Ranking, except for the host Russia, which automatically occupies the first position.
\item
The column Elo shows the strength of the teams according to the World Football Elo Ratings on 13 June 2018, see \url{https://www.international-football.net/elo-ratings-table?year=2018&month=06&day=13&confed=&}. The 2018 FIFA World Cup started on 14 June 2018. This measure quantifies the strengths of the teams in our simulation.
\end{tablenotes}
\end{threeparttable}
\end{table}

The composition of the pots is shown in Table~\ref{Table1}.

\subsection{The effect of the draw order} \label{Sec41}

According to \citet[Section~2]{RobertsRosenthal2024}, the FIFA World Cup draw procedure is unfair since some feasible allocations might occur with a higher probability.
In addition, the pre-assignment of Russia to Group A introduces a powerful bias because the draw mechanism is not independent of group labels. Russia has a 12.5\% probability of playing against an arbitrarily chosen country from Pot 2 because no draw constraints can apply. However, since there are one CONCACAF, three CONMEBOL, and four UEFA members in Pot 2, the two CONMEBOL teams from Pot 1 (Brazil and Argentina) play against a given European team from Pot 2 with a probability of 0.2 since they have five possible opponents from Pot 2. The remaining five UEFA teams in Pot 1 are identical concerning the draw constraints, thus, they have a chance of $(100 - 12.5) / 5 = 17.5$\% to be assigned to the same group as a given South American team from Pot 2. Analogously, the probability that two given European teams from Pots 1 and 2 play against each other equals $(100 - 2 \times 20 - 12.5) / 5 = 9.5$\%. 

However, an appropriate relabelling of the pots may bring the FIFA World Cup draw closer to the uniform distribution. Therefore, all possible draw orders of the pots are examined such that the pre-assignment of Russia to Group A is retained. Since the teams can be drawn in $7! \times \left( 8! \right)^3 \approx 3.3 \times 10^{17}$ different draw orders, it is almost impossible to derive exact theoretical results. Consequently, the 25 draw mechanisms---the draw procedure of FIFA with the 24 possible draw orders of the pots and the uniformly distributed rejection mechanism \citep{RobertsRosenthal2024}---are analysed on the basis of 1 million randomly generated draw orders, analogous to \citet{RobertsRosenthal2024}.

The rejection mechanism (or rejection sampler) $U$ works as follows. The groups are drawn completely randomly disregarding any geographical constraint. Then, it is checked whether all restrictions are satisfied. If yes, the draw is retained as a valid one. If no, the current draw is rejected, and a new random group draw is generated. The rejection mechanism is uniform over all possible valid draws, see, for example, \citet[Section~II.3]{Devroye1986}. In the case of the 2018 FIFA World Cup, about one of 161 random draws is accepted, which is less than one out of 560 as for the 2022 FIFA World Cup \citep{RobertsRosenthal2024}. Consequently, the total number of valid draws is approximately $7! \times \left( 8! \right)^3 / 161 \approx 2.05 \times 10^{15}$.

The distortions compared to the rejection mechanism $U$ are quantified through the difference of the probabilities that two given teams are placed in the same group. 
The bias $\Delta_{ij}$ of mechanism $M$ for teams $i$ and $j$ is suggested to be
\begin{equation*}
\Delta_{ij} = p_{ij}^M - p_{ij}^U,
\end{equation*}
where $p_{ij}^M$ and $p_{ij}^U$ are the probabilities that teams $i$ and $j$ are assigned to the same group by a draw mechanism $M$ and the rejection mechanism $U$, respectively.
When the biases are aggregated for the teams, we will always sum their \emph{absolute} value $\lvert \Delta_{ij} \rvert$.

\begin{table}[t!]
  \centering
  \caption{The deviations of different draw mechanisms in the 2018 FIFA World Cup}
  \label{Table2}
  \rowcolors{1}{}{gray!20}
\begin{threeparttable}
    \begin{tabularx}{\textwidth}{cCCLLL} \toprule
    Draw order & Minimal nonzero probability & Maximal probability & Maximum of positive ($+$) biases & Maximum of negative ($-$) biases & Sum of absolute biases \\ \bottomrule
    1-2-3-4 & 3.26\% & 32.98\% & 10.29\% (19) & 4.68\% (7) & 3.007 (1) \\
    1-2-4-3 & 1.84\% & 29.62\% & \textcolor{white}{0}9.97\% (18) & 4.18\% (1) & 3.122 (3) \\
    1-3-2-4 & 1.18\% & 28.36\% & \textcolor{gray!20}{0}8.46\% (15) & 5.17\% (11) & 3.867 (14) \\
    1-3-4-2 & 2.86\% & 28.41\% & \textcolor{white}{0}3.86\% (4) & 5.19\% (13) & 3.786 (10) \\
    1-4-2-3 & 1.05\% & 29.55\% & \textcolor{gray!20}{0}7.35\% (14) & 6.33\% (19) & 3.364 (6) \\
    1-4-3-2 & 2.12\% & 28.17\% & \textcolor{white}{0}3.21\% (1) & 6.33\% (18) & 3.756 (8) \\ \hline
    2-1-3-4 & 3.26\% & 33.29\% & 10.60\% (20) & 4.67\% (6) & 3.104 (2) \\
    2-1-4-3 & 1.75\% & 29.64\% & \textcolor{white}{0}9.96\% (17) & 4.23\% (2) & 3.206 (4) \\
    2-3-1-4 & 1.00\% & 28.44\% & \textcolor{gray!20}{0}6.08\% (9) & 5.69\% (16) & 5.174 (22) \\
    2-3-4-1 & 1.67\% & 28.62\% & \textcolor{white}{0}6.02\% (8) & 4.87\% (8) & 4.825 (21) \\
    2-4-1-3 & 1.11\% & 29.43\% & \textcolor{gray!20}{0}6.84\% (11) & 7.45\% (23) & 3.810 (11) \\
    2-4-3-1 & 1.48\% & 28.47\% & \textcolor{white}{0}6.74\% (10) & 7.53\% (24) & 4.217 (16) \\ \hline
    3-1-2-4 & 1.18\% & 28.36\% & \textcolor{gray!20}{0}8.47\% (16) & 5.17\% (12) & 3.844 (12) \\
    3-1-4-2 & 2.85\% & 28.50\% & \textcolor{white}{0}3.81\% (3) & 5.17\% (10) & 3.775 (9) \\
    3-2-1-4 & 0.95\% & 28.64\% & \textcolor{gray!20}{0}5.71\% (7) & 5.29\% (15) & 4.628 (18) \\
    3-2-4-1 & 1.37\% & 28.68\% & \textcolor{white}{0}5.48\% (6) & 5.07\% (9) & 4.397 (17) \\
    3-4-1-2 & 4.13\% & 39.07\% & 16.38\% (24) & 4.52\% (4) & 4.653 (19) \\
    3-4-2-1 & 5.72\% & 35.46\% & 12.77\% (22) & 5.26\% (14) & 5.619 (23) \\ \hline
    4-1-2-3 & 1.05\% & 29.59\% & \textcolor{gray!20}{0}7.33\% (13) & 6.27\% (17) & 3.341 (5) \\
    4-1-3-2 & 2.12\% & 28.21\% & \textcolor{white}{0}3.28\% (2) & 6.38\% (20) & 3.859 (13) \\
    4-2-1-3 & 1.09\% & 29.63\% & \textcolor{gray!20}{0}6.94\% (12) & 6.83\% (22) & 3.503 (7) \\
    4-2-3-1 & 1.35\% & 28.53\% & \textcolor{white}{0}5.24\% (5) & 6.76\% (21) & 4.138 (15) \\
    4-3-1-2 & 4.11\% & 37.96\% & 15.27\% (23) & 4.24\% (3) & 4.656 (20) \\
    4-3-2-1 & 5.47\% & 34.45\% & 11.76\% (21) & 4.67\% (5) & 5.671 (24) \\ \hline
    Uniform & 1.56\% & 28.74\% & \multicolumn{1}{c}{---}      & \multicolumn{1}{c}{---}      & \multicolumn{1}{c}{---} \\ \toprule
    \end{tabularx}
\begin{tablenotes} \footnotesize
\item
The numbers in parenthesis indicate the ranks of FIFA draw procedure with the draw order in the row according to the measure of deviation in the column.
\end{tablenotes}
\end{threeparttable}
\end{table}

The most extreme and aggregated distortions are presented in Table~\ref{Table2}. For example, the probability that Denmark (17) (or the equivalent team of Sweden (18) or Iceland (20)) and Serbia (25) play in the same group is more than doubled by the FIFA draw mechanism with the traditional draw order 1-2-3-4. On the other hand, the likelihood of assigning Russia (1) and Serbia to the same group is decreased by 4.68 percentage points, while the likelihood of assigning Mexico (14) and Serbia to the same group is increased by 10.29 percentage points.

The standard error of $\Delta_{ij}$ equals
\[
\mathit{SE}_{ij} = \sqrt{\frac{p_{ij}^M \left( 1 - p_{ij}^M \right)}{N}} + \sqrt{\frac{p_{ij}^U \left( 1 - p_{ij}^U \right)}{N}}.
\]
Since any $p_{ij}^M$ and $p_{ij}^U$ is at most $0.3907$ according to Table~\ref{Table2} and $N = 10^6$, $\mathit{SE}_{ij} \leq 0.00098$.

The last column of Table~\ref{Table2} compares the 24 draw orders by adding the values of $\Delta_{ij}$ for the 365 allowed country pairs. The bounds of the 99\% confidence intervals are given by
\[
\pm 2.58 \times \sum_{i,j} \mathit{SE}_{ij} \leq \pm 2.58 \times \sqrt{365} \times 0.00098 = \pm 0.048.
\]
Hence, the official draw order 1-2-3-4 is not only optimal but it is significantly fairer than any other draw order.
Furthermore, Table~\ref{Table2} uncovers that there are some low-performing draw orders, dominated by another draw order, such as 3-4-2-1 and 4-3-2-1.



\begin{table}[t!]
  \centering
  \caption{Fairness distortions for selected draw orders in the 2018 FIFA World Cup}
  \label{Table3}
  
\begin{subtable}{\textwidth}
  \caption{Draw order 1-2-3-4}
  \label{Table3a}
\resizebox{\textwidth}{!}{
\begin{tiny}
    \begin{tabularx}{1.4\textwidth}{r CCCC CCCC CCCC CCCC CCCC CCCC} \toprule
          & \rotatebox[origin=l]{90}{Spain} & \rotatebox[origin=l]{90}{Peru} & \rotatebox[origin=l]{90}{Switzerland} & \rotatebox[origin=l]{90}{England} & \rotatebox[origin=l]{90}{Colombia} & \rotatebox[origin=l]{90}{Mexico} & \rotatebox[origin=l]{90}{Uruguay} & \rotatebox[origin=l]{90}{Croatia} & \rotatebox[origin=l]{90}{Denmark} & \rotatebox[origin=l]{90}{Iceland} & \rotatebox[origin=l]{90}{Costa Rica} & \rotatebox[origin=l]{90}{Sweden} & \rotatebox[origin=l]{90}{Tunisia} & \rotatebox[origin=l]{90}{Egypt} & \rotatebox[origin=l]{90}{Senegal} & \rotatebox[origin=l]{90}{Iran} & \rotatebox[origin=l]{90}{Serbia} & \rotatebox[origin=l]{90}{Nigeria} & \rotatebox[origin=l]{90}{Australia} & \rotatebox[origin=l]{90}{Japan} & \rotatebox[origin=l]{90}{Morocco} & \rotatebox[origin=l]{90}{Panama} & \rotatebox[origin=l]{90}{South Korea} & \rotatebox[origin=l]{90}{Saudi Arabia} \\ \bottomrule
    Russia & \cellcolor{ForestGreen!33.144} 3.3 & \cellcolor{red!41.186} 4.1 & \cellcolor{ForestGreen!33.27} 3.3 & \cellcolor{ForestGreen!33.554} 3.4 & \cellcolor{red!42.136} 4.2 & \cellcolor{red!8.234} 0.8 & \cellcolor{red!41.741} 4.2 & \cellcolor{ForestGreen!33.329} 3.3 & \cellcolor{red!43.455} 4.3 & \cellcolor{red!43.619} 4.4 & \cellcolor{ForestGreen!11.324} 1.1 & \cellcolor{red!42.979} 4.3 & \cellcolor{ForestGreen!29.187} 2.9 & \cellcolor{ForestGreen!29.586} 3 & \cellcolor{ForestGreen!29.125} 2.9 & \cellcolor{ForestGreen!30.831} 3.1 & \cellcolor{red!46.836} 4.7 & \cellcolor{red!20.638} 2.1 & \cellcolor{ForestGreen!17.281} 1.7 & \cellcolor{ForestGreen!17.744} 1.8 & \cellcolor{red!21.536} 2.2 & \cellcolor{ForestGreen!19.345} 1.9 & \cellcolor{ForestGreen!17.31} 1.7 & \cellcolor{ForestGreen!17.33} 1.7 \\
    Germany & \cellcolor{ForestGreen!3.209} 0.3 & \cellcolor{ForestGreen!7.559} 0.8 & \cellcolor{ForestGreen!3.752} 0.4 & \cellcolor{ForestGreen!2.89} 0.3 & \cellcolor{ForestGreen!8.345} 0.8 & \cellcolor{red!37.107} 3.7 & \cellcolor{ForestGreen!7.682} 0.8 & \cellcolor{ForestGreen!3.67} 0.4 & \cellcolor{ForestGreen!4.647} 0.5 & \cellcolor{ForestGreen!4.846} 0.5 & \cellcolor{ForestGreen!1.302} 0.1 & \cellcolor{ForestGreen!5.226} 0.5 & \cellcolor{red!4.751} 0.5 & \cellcolor{red!4.086} 0.4 & \cellcolor{red!3.638} 0.4 & \cellcolor{red!3.546} 0.4 & \cellcolor{red!11.716} 1.2 & \cellcolor{ForestGreen!4.889} 0.5 & \cellcolor{red!0.281} 0 & \cellcolor{ForestGreen!0.091} 0 & \cellcolor{ForestGreen!3.371} 0.3 & \cellcolor{ForestGreen!2.082} 0.2 & \cellcolor{ForestGreen!0.869} 0.1 & \cellcolor{ForestGreen!0.695} 0.1 \\
    Brazil & \cellcolor{red!24.011} 2.4 & X     & \cellcolor{red!25.206} 2.5 & \cellcolor{red!24.447} 2.4 & X     & \cellcolor{ForestGreen!98.279} 9.8 & X     & \cellcolor{red!24.615} 2.5 & \cellcolor{ForestGreen!9.861} 1 & \cellcolor{ForestGreen!9.851} 1 & \cellcolor{red!9.612} 1 & \cellcolor{ForestGreen!8.118} 0.8 & \cellcolor{red!3.165} 0.3 & \cellcolor{red!3.836} 0.4 & \cellcolor{red!4.149} 0.4 & \cellcolor{red!7.068} 0.7 & \cellcolor{ForestGreen!52.455} 5.2 & \cellcolor{red!1.02} 0.1 & \cellcolor{red!8.401} 0.8 & \cellcolor{red!8.754} 0.9 & \cellcolor{ForestGreen!0.094} 0 & \cellcolor{red!15.948} 1.6 & \cellcolor{red!9.799} 1 & \cellcolor{red!8.627} 0.9 \\
    Portugal & \cellcolor{ForestGreen!3.534} 0.4 & \cellcolor{ForestGreen!8.829} 0.9 & \cellcolor{ForestGreen!2.918} 0.3 & \cellcolor{ForestGreen!3.593} 0.4 & \cellcolor{ForestGreen!7.552} 0.8 & \cellcolor{red!38.669} 3.9 & \cellcolor{ForestGreen!8.689} 0.9 & \cellcolor{ForestGreen!3.554} 0.4 & \cellcolor{ForestGreen!4.592} 0.5 & \cellcolor{ForestGreen!5.186} 0.5 & \cellcolor{ForestGreen!1.505} 0.2 & \cellcolor{ForestGreen!5.065} 0.5 & \cellcolor{red!4.402} 0.4 & \cellcolor{red!5.088} 0.5 & \cellcolor{red!4.191} 0.4 & \cellcolor{red!2.667} 0.3 & \cellcolor{red!11.615} 1.2 & \cellcolor{ForestGreen!3.98} 0.4 & \cellcolor{ForestGreen!0.065} 0 & \cellcolor{ForestGreen!0.269} 0 & \cellcolor{ForestGreen!4.99} 0.5 & \cellcolor{ForestGreen!2.355} 0.2 & \cellcolor{red!0.479} 0 & \cellcolor{ForestGreen!0.435} 0 \\
    Argentina & \cellcolor{red!24.155} 2.4 & X     & \cellcolor{red!24.58} 2.5 & \cellcolor{red!25.57} 2.6 & X     & \cellcolor{ForestGreen!99.137} 9.9 & X     & \cellcolor{red!24.832} 2.5 & \cellcolor{ForestGreen!10.075} 1 & \cellcolor{ForestGreen!7.944} 0.8 & \cellcolor{red!8.424} 0.8 & \cellcolor{ForestGreen!10.03} 1 & \cellcolor{red!3.699} 0.4 & \cellcolor{red!4.241} 0.4 & \cellcolor{red!4.216} 0.4 & \cellcolor{red!7.469} 0.7 & \cellcolor{ForestGreen!52.098} 5.2 & \cellcolor{ForestGreen!0.019} 0 & \cellcolor{red!9.214} 0.9 & \cellcolor{red!8.607} 0.9 & \cellcolor{red!0.48} 0 & \cellcolor{red!16.299} 1.6 & \cellcolor{red!8.698} 0.9 & \cellcolor{red!8.819} 0.9 \\
    Belgium & \cellcolor{ForestGreen!2.132} 0.2 & \cellcolor{ForestGreen!8.976} 0.9 & \cellcolor{ForestGreen!2.786} 0.3 & \cellcolor{ForestGreen!3.354} 0.3 & \cellcolor{ForestGreen!8.787} 0.9 & \cellcolor{red!37.386} 3.7 & \cellcolor{ForestGreen!8.596} 0.9 & \cellcolor{ForestGreen!2.755} 0.3 & \cellcolor{ForestGreen!5.357} 0.5 & \cellcolor{ForestGreen!5.866} 0.6 & \cellcolor{ForestGreen!0.764} 0.1 & \cellcolor{ForestGreen!4.498} 0.4 & \cellcolor{red!4.759} 0.5 & \cellcolor{red!4.83} 0.5 & \cellcolor{red!4.52} 0.5 & \cellcolor{red!2.376} 0.2 & \cellcolor{red!11.461} 1.1 & \cellcolor{ForestGreen!4.13} 0.4 & \cellcolor{red!0.168} 0 & \cellcolor{red!0.629} 0.1 & \cellcolor{ForestGreen!5.012} 0.5 & \cellcolor{ForestGreen!3.235} 0.3 & \cellcolor{red!0.224} 0 & \cellcolor{ForestGreen!0.105} 0 \\
    Poland & \cellcolor{ForestGreen!3.367} 0.3 & \cellcolor{ForestGreen!7.918} 0.8 & \cellcolor{ForestGreen!4.224} 0.4 & \cellcolor{ForestGreen!3.694} 0.4 & \cellcolor{ForestGreen!8.04} 0.8 & \cellcolor{red!38.377} 3.8 & \cellcolor{ForestGreen!8.065} 0.8 & \cellcolor{ForestGreen!3.069} 0.3 & \cellcolor{ForestGreen!4.493} 0.4 & \cellcolor{ForestGreen!4.961} 0.5 & \cellcolor{ForestGreen!1.838} 0.2 & \cellcolor{ForestGreen!4.797} 0.5 & \cellcolor{red!4.126} 0.4 & \cellcolor{red!4.126} 0.4 & \cellcolor{red!4.038} 0.4 & \cellcolor{red!3.799} 0.4 & \cellcolor{red!11.709} 1.2 & \cellcolor{ForestGreen!4.229} 0.4 & \cellcolor{red!0.186} 0 & \cellcolor{ForestGreen!0.153} 0 & \cellcolor{ForestGreen!3.839} 0.4 & \cellcolor{ForestGreen!3.019} 0.3 & \cellcolor{ForestGreen!1.484} 0.1 & \cellcolor{red!0.829} 0.1 \\
    France & \cellcolor{ForestGreen!2.78} 0.3 & \cellcolor{ForestGreen!7.904} 0.8 & \cellcolor{ForestGreen!2.836} 0.3 & \cellcolor{ForestGreen!2.932} 0.3 & \cellcolor{ForestGreen!9.412} 0.9 & \cellcolor{red!37.643} 3.8 & \cellcolor{ForestGreen!8.709} 0.9 & \cellcolor{ForestGreen!3.07} 0.3 & \cellcolor{ForestGreen!4.43} 0.4 & \cellcolor{ForestGreen!4.965} 0.5 & \cellcolor{ForestGreen!1.303} 0.1 & \cellcolor{ForestGreen!5.245} 0.5 & \cellcolor{red!4.285} 0.4 & \cellcolor{red!3.379} 0.3 & \cellcolor{red!4.373} 0.4 & \cellcolor{red!3.906} 0.4 & \cellcolor{red!11.216} 1.1 & \cellcolor{ForestGreen!4.411} 0.4 & \cellcolor{ForestGreen!0.904} 0.1 & \cellcolor{red!0.267} 0 & \cellcolor{ForestGreen!4.71} 0.5 & \cellcolor{ForestGreen!2.211} 0.2 & \cellcolor{red!0.463} 0 & \cellcolor{red!0.29} 0 \\
    Spain & X     & X     & X     & X     & X     & X     & X     & X     & \cellcolor{red!2.56} 0.3 & \cellcolor{red!4.245} 0.4 & \cellcolor{red!3.595} 0.4 & \cellcolor{red!3.259} 0.3 & \cellcolor{ForestGreen!3.115} 0.3 & \cellcolor{ForestGreen!2.931} 0.3 & \cellcolor{ForestGreen!1.896} 0.2 & \cellcolor{ForestGreen!5.717} 0.6 & \cellcolor{red!10.673} 1.1 & \cellcolor{ForestGreen!1.186} 0.1 & \cellcolor{ForestGreen!1.041} 0.1 & \cellcolor{ForestGreen!1.987} 0.2 & \cellcolor{ForestGreen!1.482} 0.1 & \cellcolor{ForestGreen!1.766} 0.2 & \cellcolor{ForestGreen!1.601} 0.2 & \cellcolor{ForestGreen!1.61} 0.2 \\
    Peru  & X     & X     & X     & X     & X     & X     & X     & X     & \cellcolor{ForestGreen!5.952} 0.6 & \cellcolor{ForestGreen!6.26} 0.6 & \cellcolor{ForestGreen!3.396} 0.3 & \cellcolor{ForestGreen!6.911} 0.7 & \cellcolor{red!4.815} 0.5 & \cellcolor{red!4.58} 0.5 & \cellcolor{red!4.877} 0.5 & \cellcolor{red!8.247} 0.8 & \cellcolor{red!20.427} 2 & \cellcolor{ForestGreen!4.198} 0.4 & \cellcolor{ForestGreen!2.478} 0.2 & \cellcolor{ForestGreen!4.172} 0.4 & \cellcolor{ForestGreen!3.764} 0.4 & \cellcolor{red!1.33} 0.1 & \cellcolor{ForestGreen!3.871} 0.4 & \cellcolor{ForestGreen!3.274} 0.3 \\
    Switzerland & X     & X     & X     & X     & X     & X     & X     & X     & \cellcolor{red!3.466} 0.3 & \cellcolor{red!3.799} 0.4 & \cellcolor{red!2.742} 0.3 & \cellcolor{red!4.26} 0.4 & \cellcolor{ForestGreen!3.044} 0.3 & \cellcolor{ForestGreen!3.138} 0.3 & \cellcolor{ForestGreen!3.498} 0.3 & \cellcolor{ForestGreen!4.587} 0.5 & \cellcolor{red!11.269} 1.1 & \cellcolor{ForestGreen!0.841} 0.1 & \cellcolor{ForestGreen!1.927} 0.2 & \cellcolor{ForestGreen!2.497} 0.2 & \cellcolor{ForestGreen!1.002} 0.1 & \cellcolor{ForestGreen!1.351} 0.1 & \cellcolor{ForestGreen!1.901} 0.2 & \cellcolor{ForestGreen!1.75} 0.2 \\
    England & X     & X     & X     & X     & X     & X     & X     & X     & \cellcolor{red!4.251} 0.4 & \cellcolor{red!2.819} 0.3 & \cellcolor{red!2.129} 0.2 & \cellcolor{red!4.345} 0.4 & \cellcolor{ForestGreen!2.819} 0.3 & \cellcolor{ForestGreen!2.407} 0.2 & \cellcolor{ForestGreen!2.124} 0.2 & \cellcolor{ForestGreen!6.194} 0.6 & \cellcolor{red!10.962} 1.1 & \cellcolor{ForestGreen!1.074} 0.1 & \cellcolor{ForestGreen!1.969} 0.2 & \cellcolor{ForestGreen!1.962} 0.2 & \cellcolor{ForestGreen!1.002} 0.1 & \cellcolor{ForestGreen!1.806} 0.2 & \cellcolor{ForestGreen!0.965} 0.1 & \cellcolor{ForestGreen!2.184} 0.2 \\
    Colombia & X     & X     & X     & X     & X     & X     & X     & X     & \cellcolor{ForestGreen!6.48} 0.6 & \cellcolor{ForestGreen!6.339} 0.6 & \cellcolor{ForestGreen!3.694} 0.4 & \cellcolor{ForestGreen!6.772} 0.7 & \cellcolor{red!5.018} 0.5 & \cellcolor{red!5.442} 0.5 & \cellcolor{red!4.404} 0.4 & \cellcolor{red!8.421} 0.8 & \cellcolor{red!19.875} 2 & \cellcolor{ForestGreen!5.094} 0.5 & \cellcolor{ForestGreen!3.048} 0.3 & \cellcolor{ForestGreen!3.125} 0.3 & \cellcolor{ForestGreen!4.848} 0.5 & \cellcolor{red!3.142} 0.3 & \cellcolor{ForestGreen!3.202} 0.3 & \cellcolor{ForestGreen!3.7} 0.4 \\
    Mexico & X     & X     & X     & X     & X     & X     & X     & X     & \cellcolor{red!4.503} 0.5 & \cellcolor{red!4.255} 0.4 & X     & \cellcolor{red!4.357} 0.4 & \cellcolor{ForestGreen!2.954} 0.3 & \cellcolor{ForestGreen!3.803} 0.4 & \cellcolor{ForestGreen!3.789} 0.4 & \cellcolor{ForestGreen!2.569} 0.3 & \cellcolor{ForestGreen!102.882} 10.3 & \cellcolor{red!16.793} 1.7 & \cellcolor{red!16.543} 1.7 & \cellcolor{red!17.237} 1.7 & \cellcolor{red!17.482} 1.7 & X     & \cellcolor{red!17.163} 1.7 & \cellcolor{red!17.664} 1.8 \\
    Uruguay & X     & X     & X     & X     & X     & X     & X     & X     & \cellcolor{ForestGreen!5.862} 0.6 & \cellcolor{ForestGreen!6.459} 0.6 & \cellcolor{ForestGreen!3.649} 0.4 & \cellcolor{ForestGreen!6.03} 0.6 & \cellcolor{red!4.947} 0.5 & \cellcolor{red!4.7} 0.5 & \cellcolor{red!4.388} 0.4 & \cellcolor{red!7.965} 0.8 & \cellcolor{red!18.605} 1.9 & \cellcolor{ForestGreen!3.952} 0.4 & \cellcolor{ForestGreen!3.696} 0.4 & \cellcolor{ForestGreen!1.836} 0.2 & \cellcolor{ForestGreen!4.522} 0.5 & \cellcolor{red!2.336} 0.2 & \cellcolor{ForestGreen!3.304} 0.3 & \cellcolor{ForestGreen!3.631} 0.4 \\
    Croatia & X     & X     & X     & X     & X     & X     & X     & X     & \cellcolor{red!3.514} 0.4 & \cellcolor{red!3.94} 0.4 & \cellcolor{red!2.273} 0.2 & \cellcolor{red!3.492} 0.3 & \cellcolor{ForestGreen!2.848} 0.3 & \cellcolor{ForestGreen!2.443} 0.2 & \cellcolor{ForestGreen!2.362} 0.2 & \cellcolor{ForestGreen!5.566} 0.6 & \cellcolor{red!11.071} 1.1 & \cellcolor{ForestGreen!0.448} 0 & \cellcolor{ForestGreen!2.384} 0.2 & \cellcolor{ForestGreen!1.658} 0.2 & \cellcolor{ForestGreen!0.862} 0.1 & \cellcolor{ForestGreen!1.885} 0.2 & \cellcolor{ForestGreen!2.319} 0.2 & \cellcolor{ForestGreen!1.515} 0.2 \\
    Denmark & \cellcolor{red!2.56} 0.3 & \cellcolor{red!4.245} 0.4 & \cellcolor{red!3.595} 0.4 & \cellcolor{red!3.259} 0.3 & \cellcolor{ForestGreen!3.115} 0.3 & \cellcolor{ForestGreen!2.931} 0.3 & \cellcolor{ForestGreen!1.896} 0.2 & \cellcolor{ForestGreen!5.717} 0.6 & X     & X     & X     & X     & X     & X     & X     & X     & \cellcolor{ForestGreen!17.109} 1.7 & \cellcolor{ForestGreen!0.992} 0.1 & \cellcolor{red!2.776} 0.3 & \cellcolor{red!3.133} 0.3 & \cellcolor{ForestGreen!0.707} 0.1 & \cellcolor{red!6.36} 0.6 & \cellcolor{red!3.321} 0.3 & \cellcolor{red!3.218} 0.3 \\
    Iceland & \cellcolor{ForestGreen!5.952} 0.6 & \cellcolor{ForestGreen!6.26} 0.6 & \cellcolor{ForestGreen!3.396} 0.3 & \cellcolor{ForestGreen!6.911} 0.7 & \cellcolor{red!4.815} 0.5 & \cellcolor{red!4.58} 0.5 & \cellcolor{red!4.877} 0.5 & \cellcolor{red!8.247} 0.8 & X     & X     & X     & X     & X     & X     & X     & X     & \cellcolor{ForestGreen!16.934} 1.7 & \cellcolor{red!0.154} 0 & \cellcolor{red!2.388} 0.2 & \cellcolor{red!3.216} 0.3 & \cellcolor{ForestGreen!0.145} 0 & \cellcolor{red!6.167} 0.6 & \cellcolor{red!2.735} 0.3 & \cellcolor{red!2.419} 0.2 \\
    Costa Rica & \cellcolor{red!3.466} 0.3 & \cellcolor{red!3.799} 0.4 & \cellcolor{red!2.742} 0.3 & \cellcolor{red!4.26} 0.4 & \cellcolor{ForestGreen!3.044} 0.3 & \cellcolor{ForestGreen!3.138} 0.3 & \cellcolor{ForestGreen!3.498} 0.3 & \cellcolor{ForestGreen!4.587} 0.5 & X     & X     & X     & X     & X     & X     & X     & X     & \cellcolor{red!5.96} 0.6 & \cellcolor{red!0.992} 0.1 & \cellcolor{ForestGreen!1.758} 0.2 & \cellcolor{ForestGreen!2.797} 0.3 & \cellcolor{red!2.01} 0.2 & X     & \cellcolor{ForestGreen!2.533} 0.3 & \cellcolor{ForestGreen!1.874} 0.2 \\
    Sweden & \cellcolor{red!4.251} 0.4 & \cellcolor{red!2.819} 0.3 & \cellcolor{red!2.129} 0.2 & \cellcolor{red!4.345} 0.4 & \cellcolor{ForestGreen!2.819} 0.3 & \cellcolor{ForestGreen!2.407} 0.2 & \cellcolor{ForestGreen!2.124} 0.2 & \cellcolor{ForestGreen!6.194} 0.6 & X     & X     & X     & X     & X     & X     & X     & X     & \cellcolor{ForestGreen!16.898} 1.7 & \cellcolor{ForestGreen!0.618} 0.1 & \cellcolor{red!2.48} 0.2 & \cellcolor{red!3.488} 0.3 & \cellcolor{ForestGreen!1.439} 0.1 & \cellcolor{red!6.752} 0.7 & \cellcolor{red!3.578} 0.4 & \cellcolor{red!2.657} 0.3 \\
    Tunisia & \cellcolor{ForestGreen!6.48} 0.6 & \cellcolor{ForestGreen!6.339} 0.6 & \cellcolor{ForestGreen!3.694} 0.4 & \cellcolor{ForestGreen!6.772} 0.7 & \cellcolor{red!5.018} 0.5 & \cellcolor{red!5.442} 0.5 & \cellcolor{red!4.404} 0.4 & \cellcolor{red!8.421} 0.8 & X     & X     & X     & X     & X     & X     & X     & X     & \cellcolor{red!7.354} 0.7 & X     & \cellcolor{ForestGreen!1.961} 0.2 & \cellcolor{ForestGreen!1.865} 0.2 & X     & \cellcolor{red!1.15} 0.1 & \cellcolor{ForestGreen!2.348} 0.2 & \cellcolor{ForestGreen!2.33} 0.2 \\
    Egypt & \cellcolor{red!4.503} 0.5 & \cellcolor{red!4.255} 0.4 & X     & \cellcolor{red!4.357} 0.4 & \cellcolor{ForestGreen!2.954} 0.3 & \cellcolor{ForestGreen!3.803} 0.4 & \cellcolor{ForestGreen!3.789} 0.4 & \cellcolor{ForestGreen!2.569} 0.3 & X     & X     & X     & X     & X     & X     & X     & X     & \cellcolor{red!7.78} 0.8 & X     & \cellcolor{ForestGreen!1.767} 0.2 & \cellcolor{ForestGreen!3.417} 0.3 & X     & \cellcolor{red!1.589} 0.2 & \cellcolor{ForestGreen!3.036} 0.3 & \cellcolor{ForestGreen!1.149} 0.1 \\
    Senegal & \cellcolor{ForestGreen!5.862} 0.6 & \cellcolor{ForestGreen!6.459} 0.6 & \cellcolor{ForestGreen!3.649} 0.4 & \cellcolor{ForestGreen!6.03} 0.6 & \cellcolor{red!4.947} 0.5 & \cellcolor{red!4.7} 0.5 & \cellcolor{red!4.388} 0.4 & \cellcolor{red!7.965} 0.8 & X     & X     & X     & X     & X     & X     & X     & X     & \cellcolor{red!7.588} 0.8 & X     & \cellcolor{ForestGreen!2.158} 0.2 & \cellcolor{ForestGreen!1.758} 0.2 & X     & \cellcolor{red!0.986} 0.1 & \cellcolor{ForestGreen!1.717} 0.2 & \cellcolor{ForestGreen!2.941} 0.3 \\
    Iran  & \cellcolor{red!3.514} 0.4 & \cellcolor{red!3.94} 0.4 & \cellcolor{red!2.273} 0.2 & \cellcolor{red!3.492} 0.3 & \cellcolor{ForestGreen!2.848} 0.3 & \cellcolor{ForestGreen!2.443} 0.2 & \cellcolor{ForestGreen!2.362} 0.2 & \cellcolor{ForestGreen!5.566} 0.6 & X     & X     & X     & X     & X     & X     & X     & X     & \cellcolor{red!22.259} 2.2 & \cellcolor{red!0.464} 0 & X     & X     & \cellcolor{red!0.281} 0 & \cellcolor{ForestGreen!23.004} 2.3 & X     & X \\ \toprule
    \end{tabularx}
\end{tiny}
}
\end{subtable}

\vspace{0.25cm}
\begin{subtable}{\textwidth}
  \caption{Draw order 4-2-3-1}
  \label{Table3b}
\resizebox{\textwidth}{!}{
\begin{threeparttable}
\begin{tiny}
    \begin{tabularx}{1.4\textwidth}{r CCCC CCCC CCCC CCCC CCCC CCCC} \toprule
          & \rotatebox[origin=l]{90}{Spain} & \rotatebox[origin=l]{90}{Peru} & \rotatebox[origin=l]{90}{Switzerland} & \rotatebox[origin=l]{90}{England} & \rotatebox[origin=l]{90}{Colombia} & \rotatebox[origin=l]{90}{Mexico} & \rotatebox[origin=l]{90}{Uruguay} & \rotatebox[origin=l]{90}{Croatia} & \rotatebox[origin=l]{90}{Denmark} & \rotatebox[origin=l]{90}{Iceland} & \rotatebox[origin=l]{90}{Costa Rica} & \rotatebox[origin=l]{90}{Sweden} & \rotatebox[origin=l]{90}{Tunisia} & \rotatebox[origin=l]{90}{Egypt} & \rotatebox[origin=l]{90}{Senegal} & \rotatebox[origin=l]{90}{Iran} & \rotatebox[origin=l]{90}{Serbia} & \rotatebox[origin=l]{90}{Nigeria} & \rotatebox[origin=l]{90}{Australia} & \rotatebox[origin=l]{90}{Japan} & \rotatebox[origin=l]{90}{Morocco} & \rotatebox[origin=l]{90}{Panama} & \rotatebox[origin=l]{90}{South Korea} & \rotatebox[origin=l]{90}{Saudi Arabia} \\ \bottomrule
    Russia & \cellcolor{ForestGreen!18.953} 1.9 & \cellcolor{red!23.66} 2.4 & \cellcolor{ForestGreen!20.093} 2 & \cellcolor{ForestGreen!20.274} 2 & \cellcolor{red!23.49} 2.3 & \cellcolor{red!8.585} 0.9 & \cellcolor{red!23.794} 2.4 & \cellcolor{ForestGreen!20.209} 2 & \cellcolor{red!36.138} 3.6 & \cellcolor{red!36.078} 3.6 & \cellcolor{ForestGreen!50.156} 5 & \cellcolor{red!36.871} 3.7 & \cellcolor{ForestGreen!16.933} 1.7 & \cellcolor{ForestGreen!17.524} 1.8 & \cellcolor{ForestGreen!17.186} 1.7 & \cellcolor{ForestGreen!7.288} 0.7 & \cellcolor{ForestGreen!21.028} 2.1 & \cellcolor{ForestGreen!0.172} 0 & \cellcolor{red!5.469} 0.5 & \cellcolor{red!4.122} 0.4 & \cellcolor{red!0.258} 0 & \cellcolor{red!2.61} 0.3 & \cellcolor{red!4.338} 0.4 & \cellcolor{red!4.403} 0.4 \\
    Germany & \cellcolor{red!9.137} 0.9 & \cellcolor{ForestGreen!5.033} 0.5 & \cellcolor{red!8.394} 0.8 & \cellcolor{red!8.843} 0.9 & \cellcolor{ForestGreen!4.449} 0.4 & \cellcolor{ForestGreen!20.863} 2.1 & \cellcolor{ForestGreen!5.085} 0.5 & \cellcolor{red!9.056} 0.9 & \cellcolor{red!7.233} 0.7 & \cellcolor{red!7.405} 0.7 & \cellcolor{ForestGreen!1.941} 0.2 & \cellcolor{red!7.751} 0.8 & \cellcolor{ForestGreen!4.33} 0.4 & \cellcolor{ForestGreen!5.734} 0.6 & \cellcolor{ForestGreen!5.551} 0.6 & \cellcolor{ForestGreen!4.833} 0.5 & \cellcolor{red!25.056} 2.5 & \cellcolor{red!0.068} 0 & \cellcolor{ForestGreen!4.74} 0.5 & \cellcolor{ForestGreen!5.177} 0.5 & \cellcolor{red!0.68} 0.1 & \cellcolor{ForestGreen!4.065} 0.4 & \cellcolor{ForestGreen!5.841} 0.6 & \cellcolor{ForestGreen!5.981} 0.6 \\
    Brazil & \cellcolor{ForestGreen!11.053} 1.1 & X     & \cellcolor{ForestGreen!11.436} 1.1 & \cellcolor{ForestGreen!12.453} 1.2 & X     & \cellcolor{red!46.389} 4.6 & X     & \cellcolor{ForestGreen!11.447} 1.1 & \cellcolor{ForestGreen!37.492} 3.7 & \cellcolor{ForestGreen!37.936} 3.8 & \cellcolor{red!29.67} 3 & \cellcolor{ForestGreen!37.981} 3.8 & \cellcolor{red!22.088} 2.2 & \cellcolor{red!22.606} 2.3 & \cellcolor{red!21.998} 2.2 & \cellcolor{red!17.047} 1.7 & \cellcolor{ForestGreen!52.405} 5.2 & \cellcolor{ForestGreen!2.396} 0.2 & \cellcolor{red!11.349} 1.1 & \cellcolor{red!11.265} 1.1 & \cellcolor{ForestGreen!1.602} 0.2 & \cellcolor{red!9.829} 1 & \cellcolor{red!12.233} 1.2 & \cellcolor{red!11.727} 1.2 \\
    Portugal & \cellcolor{red!8.288} 0.8 & \cellcolor{ForestGreen!4.335} 0.4 & \cellcolor{red!8.457} 0.8 & \cellcolor{red!7.902} 0.8 & \cellcolor{ForestGreen!3.675} 0.4 & \cellcolor{ForestGreen!19.951} 2 & \cellcolor{ForestGreen!5.04} 0.5 & \cellcolor{red!8.354} 0.8 & \cellcolor{red!8.227} 0.8 & \cellcolor{red!8.139} 0.8 & \cellcolor{ForestGreen!1.744} 0.2 & \cellcolor{red!7.777} 0.8 & \cellcolor{ForestGreen!5.952} 0.6 & \cellcolor{ForestGreen!5.26} 0.5 & \cellcolor{ForestGreen!5.357} 0.5 & \cellcolor{ForestGreen!5.83} 0.6 & \cellcolor{red!25.34} 2.5 & \cellcolor{red!1.305} 0.1 & \cellcolor{ForestGreen!5.738} 0.6 & \cellcolor{ForestGreen!5.358} 0.5 & \cellcolor{red!0.264} 0 & \cellcolor{ForestGreen!4.962} 0.5 & \cellcolor{ForestGreen!5.324} 0.5 & \cellcolor{ForestGreen!5.527} 0.6 \\
    Argentina & \cellcolor{ForestGreen!13.222} 1.3 & X     & \cellcolor{ForestGreen!10.228} 1 & \cellcolor{ForestGreen!9.842} 1 & X     & \cellcolor{red!45.034} 4.5 & X     & \cellcolor{ForestGreen!11.742} 1.2 & \cellcolor{ForestGreen!38.466} 3.8 & \cellcolor{ForestGreen!37.568} 3.8 & \cellcolor{red!29.093} 2.9 & \cellcolor{ForestGreen!38.871} 3.9 & \cellcolor{red!22.595} 2.3 & \cellcolor{red!23.274} 2.3 & \cellcolor{red!22.971} 2.3 & \cellcolor{red!16.972} 1.7 & \cellcolor{ForestGreen!51.89} 5.2 & \cellcolor{ForestGreen!1.689} 0.2 & \cellcolor{red!11.469} 1.1 & \cellcolor{red!11.7} 1.2 & \cellcolor{ForestGreen!1.787} 0.2 & \cellcolor{red!9.949} 1 & \cellcolor{red!11.002} 1.1 & \cellcolor{red!11.246} 1.1 \\
    Belgium & \cellcolor{red!9.425} 0.9 & \cellcolor{ForestGreen!5.294} 0.5 & \cellcolor{red!8.226} 0.8 & \cellcolor{red!8.136} 0.8 & \cellcolor{ForestGreen!4.723} 0.5 & \cellcolor{ForestGreen!20.408} 2 & \cellcolor{ForestGreen!4.205} 0.4 & \cellcolor{red!8.843} 0.9 & \cellcolor{red!8.355} 0.8 & \cellcolor{red!7.145} 0.7 & \cellcolor{ForestGreen!1.73} 0.2 & \cellcolor{red!7.88} 0.8 & \cellcolor{ForestGreen!5.837} 0.6 & \cellcolor{ForestGreen!5.333} 0.5 & \cellcolor{ForestGreen!5.146} 0.5 & \cellcolor{ForestGreen!5.334} 0.5 & \cellcolor{red!24.903} 2.5 & \cellcolor{red!0.789} 0.1 & \cellcolor{ForestGreen!5.642} 0.6 & \cellcolor{ForestGreen!4.982} 0.5 & \cellcolor{red!0.93} 0.1 & \cellcolor{ForestGreen!4.216} 0.4 & \cellcolor{ForestGreen!5.976} 0.6 & \cellcolor{ForestGreen!5.806} 0.6 \\
    Poland & \cellcolor{red!8.075} 0.8 & \cellcolor{ForestGreen!4.458} 0.4 & \cellcolor{red!7.614} 0.8 & \cellcolor{red!8.735} 0.9 & \cellcolor{ForestGreen!4.512} 0.5 & \cellcolor{ForestGreen!19.553} 2 & \cellcolor{ForestGreen!4.644} 0.5 & \cellcolor{red!8.743} 0.9 & \cellcolor{red!8.081} 0.8 & \cellcolor{red!8.319} 0.8 & \cellcolor{ForestGreen!1.557} 0.2 & \cellcolor{red!8.074} 0.8 & \cellcolor{ForestGreen!6.032} 0.6 & \cellcolor{ForestGreen!6.078} 0.6 & \cellcolor{ForestGreen!5.506} 0.6 & \cellcolor{ForestGreen!5.301} 0.5 & \cellcolor{red!25.064} 2.5 & \cellcolor{red!1.094} 0.1 & \cellcolor{ForestGreen!5.246} 0.5 & \cellcolor{ForestGreen!6.045} 0.6 & \cellcolor{red!0.942} 0.1 & \cellcolor{ForestGreen!4.748} 0.5 & \cellcolor{ForestGreen!5.615} 0.6 & \cellcolor{ForestGreen!5.446} 0.5 \\
    France & \cellcolor{red!8.303} 0.8 & \cellcolor{ForestGreen!4.54} 0.5 & \cellcolor{red!9.066} 0.9 & \cellcolor{red!8.953} 0.9 & \cellcolor{ForestGreen!6.131} 0.6 & \cellcolor{ForestGreen!19.233} 1.9 & \cellcolor{ForestGreen!4.82} 0.5 & \cellcolor{red!8.402} 0.8 & \cellcolor{red!7.924} 0.8 & \cellcolor{red!8.418} 0.8 & \cellcolor{ForestGreen!1.635} 0.2 & \cellcolor{red!8.499} 0.8 & \cellcolor{ForestGreen!5.599} 0.6 & \cellcolor{ForestGreen!5.951} 0.6 & \cellcolor{ForestGreen!6.223} 0.6 & \cellcolor{ForestGreen!5.433} 0.5 & \cellcolor{red!24.96} 2.5 & \cellcolor{red!1.001} 0.1 & \cellcolor{ForestGreen!6.921} 0.7 & \cellcolor{ForestGreen!5.525} 0.6 & \cellcolor{red!0.315} 0 & \cellcolor{ForestGreen!4.397} 0.4 & \cellcolor{ForestGreen!4.817} 0.5 & \cellcolor{ForestGreen!4.616} 0.5 \\
    Spain & X     & X     & X     & X     & X     & X     & X     & X     & \cellcolor{ForestGreen!24.911} 2.5 & \cellcolor{ForestGreen!23.715} 2.4 & \cellcolor{red!13.867} 1.4 & \cellcolor{ForestGreen!25.6} 2.6 & \cellcolor{red!15.97} 1.6 & \cellcolor{red!16.815} 1.7 & \cellcolor{red!16.414} 1.6 & \cellcolor{red!11.16} 1.1 & \cellcolor{ForestGreen!34.249} 3.4 & \cellcolor{ForestGreen!5.599} 0.6 & \cellcolor{red!9.103} 0.9 & \cellcolor{red!9.916} 1 & \cellcolor{ForestGreen!4.882} 0.5 & \cellcolor{red!8.456} 0.8 & \cellcolor{red!8.63} 0.9 & \cellcolor{red!8.625} 0.9 \\
    Peru  & X     & X     & X     & X     & X     & X     & X     & X     & \cellcolor{red!25.521} 2.6 & \cellcolor{red!25.249} 2.5 & \cellcolor{ForestGreen!17.507} 1.8 & \cellcolor{red!26.112} 2.6 & \cellcolor{ForestGreen!15.787} 1.6 & \cellcolor{ForestGreen!15.774} 1.6 & \cellcolor{ForestGreen!15.227} 1.5 & \cellcolor{ForestGreen!12.587} 1.3 & \cellcolor{red!23.352} 2.3 & \cellcolor{red!6.052} 0.6 & \cellcolor{ForestGreen!6.292} 0.6 & \cellcolor{ForestGreen!6.76} 0.7 & \cellcolor{red!7.554} 0.8 & \cellcolor{ForestGreen!11.394} 1.1 & \cellcolor{ForestGreen!6.422} 0.6 & \cellcolor{ForestGreen!6.09} 0.6 \\
    Switzerland & X     & X     & X     & X     & X     & X     & X     & X     & \cellcolor{ForestGreen!24.264} 2.4 & \cellcolor{ForestGreen!24.76} 2.5 & \cellcolor{red!13.245} 1.3 & \cellcolor{ForestGreen!24.074} 2.4 & \cellcolor{red!16.179} 1.6 & \cellcolor{red!15.965} 1.6 & \cellcolor{red!15.968} 1.6 & \cellcolor{red!11.741} 1.2 & \cellcolor{ForestGreen!33.667} 3.4 & \cellcolor{ForestGreen!4.675} 0.5 & \cellcolor{red!8.584} 0.9 & \cellcolor{red!8.11} 0.8 & \cellcolor{ForestGreen!5.077} 0.5 & \cellcolor{red!8.896} 0.9 & \cellcolor{red!8.902} 0.9 & \cellcolor{red!8.927} 0.9 \\
    England & X     & X     & X     & X     & X     & X     & X     & X     & \cellcolor{ForestGreen!23.942} 2.4 & \cellcolor{ForestGreen!24.954} 2.5 & \cellcolor{red!13.834} 1.4 & \cellcolor{ForestGreen!23.848} 2.4 & \cellcolor{red!15.925} 1.6 & \cellcolor{red!16.05} 1.6 & \cellcolor{red!15.821} 1.6 & \cellcolor{red!11.114} 1.1 & \cellcolor{ForestGreen!34.296} 3.4 & \cellcolor{ForestGreen!4.74} 0.5 & \cellcolor{red!8.648} 0.9 & \cellcolor{red!9.079} 0.9 & \cellcolor{ForestGreen!4.76} 0.5 & \cellcolor{red!8.476} 0.8 & \cellcolor{red!8.968} 0.9 & \cellcolor{red!8.625} 0.9 \\
    Colombia & X     & X     & X     & X     & X     & X     & X     & X     & \cellcolor{red!25.752} 2.6 & \cellcolor{red!26.082} 2.6 & \cellcolor{ForestGreen!18.381} 1.8 & \cellcolor{red!26.445} 2.6 & \cellcolor{ForestGreen!16.176} 1.6 & \cellcolor{ForestGreen!15.616} 1.6 & \cellcolor{ForestGreen!15.49} 1.5 & \cellcolor{ForestGreen!12.616} 1.3 & \cellcolor{red!22.899} 2.3 & \cellcolor{red!6.591} 0.7 & \cellcolor{ForestGreen!6.147} 0.6 & \cellcolor{ForestGreen!6.765} 0.7 & \cellcolor{red!7.126} 0.7 & \cellcolor{ForestGreen!11.616} 1.2 & \cellcolor{ForestGreen!5.993} 0.6 & \cellcolor{ForestGreen!6.095} 0.6 \\
    Mexico & X     & X     & X     & X     & X     & X     & X     & X     & \cellcolor{red!20.623} 2.1 & \cellcolor{red!20.762} 2.1 & X     & \cellcolor{red!19.688} 2 & \cellcolor{ForestGreen!17.319} 1.7 & \cellcolor{ForestGreen!17.84} 1.8 & \cellcolor{ForestGreen!18.016} 1.8 & \cellcolor{ForestGreen!7.898} 0.8 & \cellcolor{red!67.64} 6.8 & \cellcolor{ForestGreen!0.194} 0 & \cellcolor{ForestGreen!16.63} 1.7 & \cellcolor{ForestGreen!15.872} 1.6 & \cellcolor{ForestGreen!1.344} 0.1 & X     & \cellcolor{ForestGreen!16.449} 1.6 & \cellcolor{ForestGreen!17.151} 1.7 \\
    Uruguay & X     & X     & X     & X     & X     & X     & X     & X     & \cellcolor{red!25.849} 2.6 & \cellcolor{red!25.683} 2.6 & \cellcolor{ForestGreen!18.656} 1.9 & \cellcolor{red!26.252} 2.6 & \cellcolor{ForestGreen!15.337} 1.5 & \cellcolor{ForestGreen!15.617} 1.6 & \cellcolor{ForestGreen!16.254} 1.6 & \cellcolor{ForestGreen!11.92} 1.2 & \cellcolor{red!22.391} 2.2 & \cellcolor{red!7.422} 0.7 & \cellcolor{ForestGreen!6.695} 0.7 & \cellcolor{ForestGreen!6.135} 0.6 & \cellcolor{red!6.073} 0.6 & \cellcolor{ForestGreen!11.206} 1.1 & \cellcolor{ForestGreen!5.868} 0.6 & \cellcolor{ForestGreen!5.982} 0.6 \\
    Croatia & X     & X     & X     & X     & X     & X     & X     & X     & \cellcolor{ForestGreen!24.628} 2.5 & \cellcolor{ForestGreen!24.347} 2.4 & \cellcolor{red!13.598} 1.4 & \cellcolor{ForestGreen!24.975} 2.5 & \cellcolor{red!16.545} 1.7 & \cellcolor{red!16.017} 1.6 & \cellcolor{red!16.784} 1.7 & \cellcolor{red!11.006} 1.1 & \cellcolor{ForestGreen!34.07} 3.4 & \cellcolor{ForestGreen!4.857} 0.5 & \cellcolor{red!9.429} 0.9 & \cellcolor{red!8.427} 0.8 & \cellcolor{ForestGreen!4.69} 0.5 & \cellcolor{red!8.388} 0.8 & \cellcolor{red!8.232} 0.8 & \cellcolor{red!9.141} 0.9 \\
    Denmark & \cellcolor{ForestGreen!24.911} 2.5 & \cellcolor{ForestGreen!23.715} 2.4 & \cellcolor{red!13.867} 1.4 & \cellcolor{ForestGreen!25.6} 2.6 & \cellcolor{red!15.97} 1.6 & \cellcolor{red!16.815} 1.7 & \cellcolor{red!16.414} 1.6 & \cellcolor{red!11.16} 1.1 & X     & X     & X     & X     & X     & X     & X     & X     & \cellcolor{red!2.014} 0.2 & \cellcolor{red!2.151} 0.2 & \cellcolor{ForestGreen!0.803} 0.1 & \cellcolor{ForestGreen!1.501} 0.2 & \cellcolor{red!2.177} 0.2 & \cellcolor{ForestGreen!2.667} 0.3 & \cellcolor{ForestGreen!0.313} 0 & \cellcolor{ForestGreen!1.058} 0.1 \\
    Iceland & \cellcolor{red!25.521} 2.6 & \cellcolor{red!25.249} 2.5 & \cellcolor{ForestGreen!17.507} 1.8 & \cellcolor{red!26.112} 2.6 & \cellcolor{ForestGreen!15.787} 1.6 & \cellcolor{ForestGreen!15.774} 1.6 & \cellcolor{ForestGreen!15.227} 1.5 & \cellcolor{ForestGreen!12.587} 1.3 & X     & X     & X     & X     & X     & X     & X     & X     & \cellcolor{red!2.14} 0.2 & \cellcolor{red!2.682} 0.3 & \cellcolor{ForestGreen!0.969} 0.1 & \cellcolor{ForestGreen!0.434} 0 & \cellcolor{red!2.071} 0.2 & \cellcolor{ForestGreen!3.165} 0.3 & \cellcolor{ForestGreen!1.615} 0.2 & \cellcolor{ForestGreen!0.71} 0.1 \\
    Costa Rica & \cellcolor{ForestGreen!24.264} 2.4 & \cellcolor{ForestGreen!24.76} 2.5 & \cellcolor{red!13.245} 1.3 & \cellcolor{ForestGreen!24.074} 2.4 & \cellcolor{red!16.179} 1.6 & \cellcolor{red!15.965} 1.6 & \cellcolor{red!15.968} 1.6 & \cellcolor{red!11.741} 1.2 & X     & X     & X     & X     & X     & X     & X     & X     & \cellcolor{ForestGreen!7.135} 0.7 & \cellcolor{ForestGreen!9.41} 0.9 & \cellcolor{red!7.058} 0.7 & \cellcolor{red!5.254} 0.5 & \cellcolor{ForestGreen!8.813} 0.9 & X     & \cellcolor{red!5.999} 0.6 & \cellcolor{red!7.047} 0.7 \\
    Sweden & \cellcolor{ForestGreen!23.942} 2.4 & \cellcolor{ForestGreen!24.954} 2.5 & \cellcolor{red!13.834} 1.4 & \cellcolor{ForestGreen!23.848} 2.4 & \cellcolor{red!15.925} 1.6 & \cellcolor{red!16.05} 1.6 & \cellcolor{red!15.821} 1.6 & \cellcolor{red!11.114} 1.1 & X     & X     & X     & X     & X     & X     & X     & X     & \cellcolor{red!2.114} 0.2 & \cellcolor{red!2.485} 0.2 & \cellcolor{ForestGreen!1.777} 0.2 & \cellcolor{ForestGreen!0.717} 0.1 & \cellcolor{red!1.992} 0.2 & \cellcolor{ForestGreen!2.888} 0.3 & \cellcolor{red!0.094} 0 & \cellcolor{ForestGreen!1.303} 0.1 \\
    Tunisia & \cellcolor{red!25.752} 2.6 & \cellcolor{red!26.082} 2.6 & \cellcolor{ForestGreen!18.381} 1.8 & \cellcolor{red!26.445} 2.6 & \cellcolor{ForestGreen!16.176} 1.6 & \cellcolor{ForestGreen!15.616} 1.6 & \cellcolor{ForestGreen!15.49} 1.5 & \cellcolor{ForestGreen!12.616} 1.3 & X     & X     & X     & X     & X     & X     & X     & X     & \cellcolor{red!1.279} 0.1 & X     & \cellcolor{ForestGreen!1.685} 0.2 & \cellcolor{ForestGreen!0.399} 0 & X     & \cellcolor{red!3.155} 0.3 & \cellcolor{ForestGreen!1.364} 0.1 & \cellcolor{ForestGreen!0.986} 0.1 \\
    Egypt & \cellcolor{red!20.623} 2.1 & \cellcolor{red!20.762} 2.1 & X     & \cellcolor{red!19.688} 2 & \cellcolor{ForestGreen!17.319} 1.7 & \cellcolor{ForestGreen!17.84} 1.8 & \cellcolor{ForestGreen!18.016} 1.8 & \cellcolor{ForestGreen!7.898} 0.8 & X     & X     & X     & X     & X     & X     & X     & X     & \cellcolor{red!1.438} 0.1 & X     & \cellcolor{ForestGreen!0.525} 0.1 & \cellcolor{ForestGreen!1.102} 0.1 & X     & \cellcolor{red!3.631} 0.4 & \cellcolor{ForestGreen!1.85} 0.2 & \cellcolor{ForestGreen!1.592} 0.2 \\
    Senegal & \cellcolor{red!25.849} 2.6 & \cellcolor{red!25.683} 2.6 & \cellcolor{ForestGreen!18.656} 1.9 & \cellcolor{red!26.252} 2.6 & \cellcolor{ForestGreen!15.337} 1.5 & \cellcolor{ForestGreen!15.617} 1.6 & \cellcolor{ForestGreen!16.254} 1.6 & \cellcolor{ForestGreen!11.92} 1.2 & X     & X     & X     & X     & X     & X     & X     & X     & \cellcolor{red!1.42} 0.1 & X     & \cellcolor{ForestGreen!1.299} 0.1 & \cellcolor{ForestGreen!1.101} 0.1 & X     & \cellcolor{red!3.329} 0.3 & \cellcolor{ForestGreen!0.951} 0.1 & \cellcolor{ForestGreen!1.398} 0.1 \\
    Iran  & \cellcolor{ForestGreen!24.628} 2.5 & \cellcolor{ForestGreen!24.347} 2.4 & \cellcolor{red!13.598} 1.4 & \cellcolor{ForestGreen!24.975} 2.5 & \cellcolor{red!16.545} 1.7 & \cellcolor{red!16.017} 1.6 & \cellcolor{red!16.784} 1.7 & \cellcolor{red!11.006} 1.1 & X     & X     & X     & X     & X     & X     & X     & X     & \cellcolor{ForestGreen!3.27} 0.3 & \cellcolor{red!2.092} 0.2 & X     & X     & \cellcolor{red!2.573} 0.3 & \cellcolor{ForestGreen!1.395} 0.1 & X     & X \\ \toprule
    \end{tabularx}
\end{tiny}
    \begin{tablenotes} \footnotesize
\item
X represents a pair of teams that cannot play in the group stage.
\item
The numbers show percentages ($100 \times \Delta_{ij}$) rounded to one decimal place.
\item
\textcolor{ForestGreen}{Green} (\textcolor{red}{Red}) colour means that the FIFA draw procedure implies a higher (lower) probability than a uniform draw. For instance, the probability of assigning Mexico and Serbia to the same group is higher (lower) by 10.3 (6.8) percentage points under the FIFA draw mechanism with the draw order 1-2-3-4 (draw order 4-2-3-1) than under a uniform draw when it is 22.7\%.
\item
Darker colour indicates a higher value.
    \end{tablenotes}
\end{threeparttable}
}
\end{subtable}
\end{table}


Table~\ref{Table3} presents fairness distortions for two fundamentally different draw orders. Clearly, Russia and Mexico pose challenges for both of them. However, the official mechanism (draw order 1-2-3-4) is clearly better for the country pairs from Pot 1 and Pot 2, as well as from Pot 1 and Pot 3, which is favourable as these teams have the highest chance to qualify for the Round of 16.

The distortions are also worth studying by taking the average of absolute or squared biases for all country pairs that involve a given national team.
For the sake of simplicity, four draw orders are analysed in detail: the official 1-2-3-4, one that improves both the minimal nonzero and maximal probabilities with a slight change (1-2-4-3), one that is less biased for Mexico (1-3-2-4), and a fundamentally different draw order (4-2-3-1).

\begin{figure}[t!]
\centering

\begin{tikzpicture}
\begin{axis}[
name = axis1,
width = 0.5\textwidth, 
height = 0.8\textwidth,
title = {Mean absolute deviations},
title style = {align=center, font=\small},
xmajorgrids = true,
ymajorgrids = true,
xmin = 0,
scaled x ticks = false,
xlabel = {Mean absolute deviation},
xlabel style = {align=center, font=\small},
xticklabel style = {/pgf/number format/fixed,/pgf/number format/precision=5},
ytick style = {draw = none},
ymin = 1,
ymax = 32,
y dir = reverse,
ylabel = {Rank of the national team in the ranking used for the draw},
ylabel style = {align=center, font=\small},
legend style = {font=\small,at={(0.3,-0.15)},anchor=north west,legend columns=2},
legend entries = {Official (draw order 1-2-3-4)$\qquad$, Draw order 1-2-4-3, Draw order 1-3-2-4$\qquad \qquad \qquad$,Draw order 4-2-3-1},
]
\addplot [red, only marks, mark = star, thick] coordinates{
(0.0293633333333333,1)
(0.00542708333333333,2)
(0.0170150476190476,3)
(0.00559258333333333,4)
(0.0170764761904762,5)
(0.00552941666666667,6)
(0.00559916666666667,7)
(0.00548516666666667,8)
(0.0060415,9)
(0.00776927272727273,10)
(0.00627683333333333,11)
(0.00621025,12)
(0.00803981818181818,13)
(0.0284920909090909,14)
(0.00769836363636364,15)
(0.00614475,16)
(0.00671308333333333,17)
(0.00664633333333333,18)
(0.00343063636363636,19)
(0.00680391666666667,20)
(0.00477009090909091,21)
(0.00487972727272727,22)
(0.00466981818181818,23)
(0.0078468,24)
(0.0215313333333333,25)
(0.00381533333333333,26)
(0.00369017391304348,27)
(0.00394182608695652,28)
(0.00397990476190476,29)
(0.00564172727272727,30)
(0.00404,31)
(0.0038715652173913,32)
};
\addplot [blue, only marks, mark = square, very thick] coordinates{
(0.02328025,1)
(0.00476016666666667,2)
(0.0172477142857143,3)
(0.00484241666666667,4)
(0.0175537142857143,5)
(0.00489916666666667,6)
(0.004865,7)
(0.00488483333333333,8)
(0.00738825,9)
(0.00758227272727273,10)
(0.00748908333333333,11)
(0.00763666666666667,12)
(0.00728972727272727,13)
(0.0281981818181818,14)
(0.00746218181818182,15)
(0.00754275,16)
(0.00842783333333333,17)
(0.00836166666666667,18)
(0.0158235454545455,19)
(0.00835766666666667,20)
(0.00407372727272727,21)
(0.00407590909090909,22)
(0.004107,23)
(0.0071527,24)
(0.00804616666666667,25)
(0.0101859047619048,26)
(0.00483026086956522,27)
(0.00455252173913044,28)
(0.0102346666666667,29)
(0.00544309090909091,30)
(0.00469547826086956,31)
(0.00475260869565217,32)
};
\addplot [brown, only marks, mark = pentagon, very thick] coordinates{
(0.0138394166666667,1)
(0.00742883333333333,2)
(0.0235724761904762,3)
(0.00733,4)
(0.0236471428571429,5)
(0.00734175,6)
(0.00744766666666667,7)
(0.00729925,8)
(0.00920433333333333,9)
(0.0116480909090909,10)
(0.0090385,11)
(0.00907566666666667,12)
(0.0115541818181818,13)
(0.00606481818181818,14)
(0.0114934545454545,15)
(0.0090685,16)
(0.0164978333333333,17)
(0.0165488333333333,18)
(0.00995927272727273,19)
(0.0163546666666667,20)
(0.0110194545454545,21)
(0.0109110909090909,22)
(0.0111300909090909,23)
(0.0140871,24)
(0.02613175,25)
(0.00721209523809524,26)
(0.00369626086956522,27)
(0.00362113043478261,28)
(0.00719952380952381,29)
(0.00373663636363636,30)
(0.00348686956521739,31)
(0.00357826086956522,32)
};
\addplot [ForestGreen, only marks, mark = triangle, very thick] coordinates{
(0.0174846666666667,1)
(0.00696858333333333,2)
(0.0205905714285714,3)
(0.00700441666666667,4)
(0.0205052380952381,5)
(0.00705266666666667,6)
(0.00706175,7)
(0.00715341666666667,8)
(0.0135153333333333,9)
(0.0125,10)
(0.0131895,11)
(0.0132590833333333,12)
(0.0125804545454545,13)
(0.0207928181818182,14)
(0.0124967272727273,15)
(0.01341375,16)
(0.01500375,17)
(0.0150144166666667,18)
(0.0126059090909091,19)
(0.0151695,20)
(0.0103396363636364,21)
(0.0105269090909091,22)
(0.0104277272727273,23)
(0.0083705,24)
(0.0226675,25)
(0.00321257142857143,26)
(0.00618339130434783,27)
(0.005902,28)
(0.00313857142857143,29)
(0.00606536363636364,30)
(0.00594765217391304,31)
(0.0060644347826087,32)
};
\end{axis}

\begin{axis}[
at = {(axis1.south east)},
xshift = 0.1\textwidth,
width = 0.5\textwidth, 
height = 0.8\textwidth,
title = {Mean squared deviations},
title style = {align=center, font=\small},
xmajorgrids = true,
ymajorgrids = true,
xmode = log,
xlabel = {Mean squared deviation \\ (logarithmic scale)},
scaled x ticks = false,
xlabel style = {align=center, font=\small},
xticklabel style = {/pgf/number format/fixed,/pgf/number format/precision=4},
ytick style = {draw = none},
ymin = 1,
ymax = 32,
ylabel = {Rank of the national team in the ranking used for the draw},
ylabel style = {align=center, font=\small},
y dir = reverse,
]
\addplot [red, only marks, mark = star, thick] coordinates{
(0.000986973901416667,1)
(8.02611748333334E-05,2)
(0.000754439675238095,3)
(8.66604875833333E-05,4)
(0.000762863953809524,5)
(8.33745181666667E-05,6)
(8.47990194166667E-05,7)
(0.000082868828,8)
(0.000105767817083333,9)
(0.000127593279636364,10)
(0.000110595983,11)
(0.000111548493583333,12)
(0.000132576113,13)
(0.00178025641909091,14)
(0.000126155593181818,15)
(0.000109371781416667,16)
(0.000114703040666667,17)
(0.000114422854083333,18)
(1.96013721818182E-05,19)
(0.000113249444083333,20)
(0.000052991984,21)
(5.49145091818182E-05,22)
(0.000052296394,23)
(0.0001234147002,24)
(0.0009216968635,25)
(4.13505286666667E-05,26)
(3.50636833913043E-05,27)
(3.77810312173913E-05,28)
(4.46543846666667E-05,29)
(7.34320631818182E-05,30)
(3.79729327826087E-05,31)
(3.73256700869565E-05,32)
};
\addplot [blue, only marks, mark = square, very thick] coordinates{
(0.0007463756495,1)
(7.53608348333333E-05,2)
(0.000682385686095238,3)
(0.00007909846325,4)
(0.000707418061714286,5)
(0.00007771495125,6)
(0.00007603089775,7)
(0.00008152035875,8)
(0.000119468055,9)
(0.000129794550636364,10)
(0.000124141086166667,11)
(0.000127716570666667,12)
(0.000125120740545455,13)
(0.00141575216363636,14)
(0.000126574782454545,15)
(0.000125407806333333,16)
(0.000145547979833333,17)
(0.000144417046333333,18)
(0.000422936476363636,19)
(0.000144636266166667,20)
(4.17156240909091E-05,21)
(4.07686426363636E-05,22)
(4.14664011818182E-05,23)
(0.0000760465553,24)
(0.000136550555083333,25)
(0.000180409765714286,26)
(3.37324817391304E-05,27)
(2.86012756521739E-05,28)
(0.000177443563238095,29)
(3.82276361818182E-05,30)
(3.14101691304348E-05,31)
(0.000031941252173913,32)
};
\addplot [brown, only marks, mark = pentagon, very thick] coordinates{
(0.000256136639083333,1)
(9.49963510833333E-05,2)
(0.001002125194,3)
(9.32746643333334E-05,4)
(0.000995837613428571,5)
(9.65546525833333E-05,6)
(9.54658143333333E-05,7)
(9.36580544166667E-05,8)
(0.000143421042583333,9)
(0.000245153790272727,10)
(0.000140324235166667,11)
(0.000143848437833333,12)
(0.000243416883818182,13)
(5.73047290909091E-05,14)
(0.000240759950636364,15)
(0.000140760144083333,16)
(0.000466119493416667,17)
(0.0004671481,18)
(0.000160452728636364,19)
(0.00045765438075,20)
(0.000190075258909091,21)
(0.000186667395454545,22)
(0.000191863599363636,23)
(0.000245933406,24)
(0.001162517548,25)
(6.90621102857143E-05,26)
(2.90374475652174E-05,27)
(2.95651586086957E-05,28)
(6.91761834285714E-05,29)
(3.38588428181818E-05,30)
(2.92160904347826E-05,31)
(2.89741187826087E-05,32)
};
\addplot [ForestGreen, only marks, mark = triangle, very thick] coordinates{
(0.000470540384,1)
(7.75478488333333E-05,2)
(0.000620292818476191,3)
(7.66854254166667E-05,4)
(0.000617716810476191,5)
(7.72615569166667E-05,6)
(7.63326925833333E-05,7)
(0.0000776317425,8)
(0.000234310870083333,9)
(0.000216835640909091,10)
(0.00022557594825,11)
(0.000228022705,12)
(0.000220942158363636,13)
(0.000644322537727273,14)
(0.000218266657,15)
(0.000232641079333333,16)
(0.00038878375175,17)
(0.000387251229166667,18)
(0.000289720003636364,19)
(0.000397781006833333,20)
(0.000161306732090909,21)
(0.000165929576181818,22)
(0.000163536460818182,23)
(0.0000913963316,24)
(0.000828849614333333,25)
(1.70726677142857E-05,26)
(5.35326265217391E-05,27)
(5.04298914782609E-05,28)
(0.000016489984,29)
(4.76368859090909E-05,30)
(0.000050929566173913,31)
(5.25664699130435E-05,32)
};
\end{axis}
\end{tikzpicture}

\caption{The average biases of different draw procedures \\ for the national teams in the 2018 FIFA World Cup}
\label{Fig3}

\end{figure}
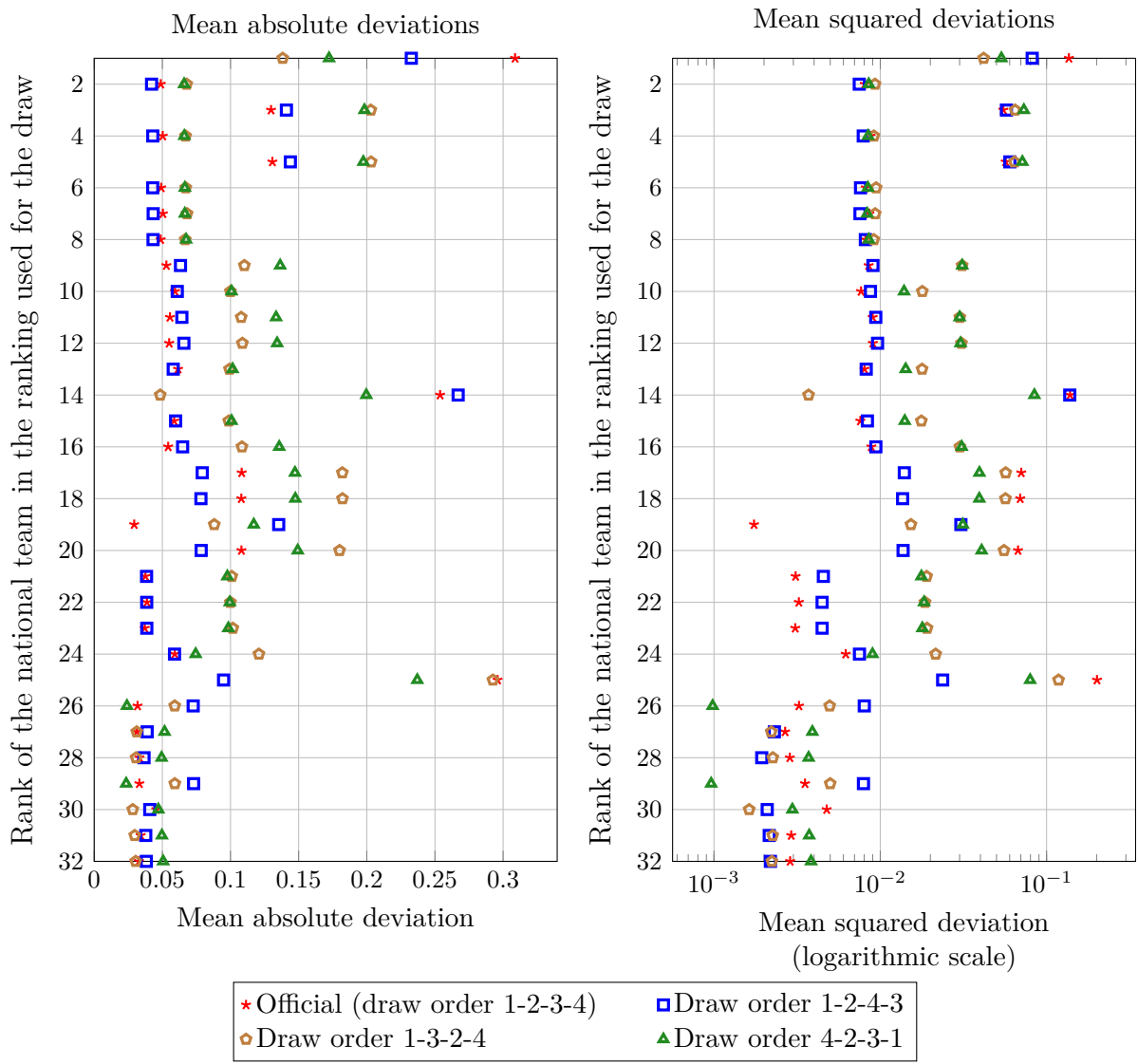


Figure~\ref{Fig3} compares the means of the biases for the 32 countries. The official draw procedure is far from the fair rejection mechanism $U$ in the case of Russia (1), Brazil (3), Argentina (5), Mexico (14), and Serbia (25). Note that Brazil and Argentina are interchangeable in the draw, thus, the deviations for these two countries are different only due to the (seemingly low) sampling error.
Reversing the draw order of Pots 3 and 4 reduces unfairness for Russia and Serbia but it does not treat the problem for Mexico. Contrarily, the deviation for Costa Rica (19) strongly increases with the draw order 1-2-4-3. Drawing Pot 3 immediately after the best teams in Pot 1 implies a relatively low distortion in the case of Russia, and, especially, Mexico but the situation of the African (21, 22, 23) and European (17, 18, 20) teams in Pot 3 becomes less fair compared to the draw order 1-2-4-3. Finally, the draw order 4-2-3-1 is unfair with respect to all countries in Pot 2 except for Mexico.

\subsection{The consequences of non-uniform distribution} \label{Sec42}

In the 2018 FIFA World Cup, the top two teams from each group have advanced to the Round of 16. Therefore, the distortions of the draw procedure are important primarily if they affect the probability of qualification for the knockout stage. To that end, the simulation methodology of \citet{FootballRankings2020} is used. This models the number of goals scored in a match by Poisson distribution: the expected number of goals is a quartic polynomial of win expectancy as estimated by a least squares regression based on more than 29 thousand home-away games and almost 10 thousand games played on neutral ground between national football teams \citep{FootballRankings2020}. Win expectancy depends on the strengths of the teams according to a well-established metric \citep{LasekSzlavikBhulai2013, GasquezRoyuela2016}, the World Football Elo ratings (\href{http://eloratings.net/about}{http://eloratings.net/about}), and the field of the match, which is neutral except for Russia, the host.

In particular, the probability that team $i$ scores $k$ goals against team $j$ equals
\begin{equation*} \label{Poisson_dist}
P_{ij}(k) = \frac{ \left( \lambda_{ij}^{(f)} \right)^k \exp \left( -\lambda_{ij}^{(f)} \right)}{k!},
\end{equation*}
where $\lambda_{ij}^{(f)}$ is the expected number of goals scored by team $i$ against team $j$ if the match is played on field $f$ (home: $f = h$; away: $f = a$; neutral: $f = n$).

World Football Elo ratings determine the win expectancy $W_{ij}$ of team $i$ against team $j$ as
\begin{equation*} \label{eq1}
W_{ij} = \frac{1}{1 + 10^{-(E_i - E_j)/400}},
\end{equation*}
where $E_i$ and $E_j$ are the Elo ratings of the two teams, respectively. The rating of the home team (here, Russia) is increased by $100$ to reflect home advantage.

\citet{FootballRankings2020} estimates how $\lambda_{ij}^{(f)}$ depends on $W_{ij}$ by least squares regressions with a regime change at $W_{ij} = 0.9$ due to the excessive number of goals scored in unbalanced matches.
Most games of the FIFA World Cup are played on neutral field when
\begin{equation*} \label{Exp_goals_neutral}
\lambda_{ij}^{(n)} = 
\left\{ \begin{array}{ll}
3.90388 \cdot W_{ij}^4 - 0.58486 \cdot W_{ij}^3 \\
- 2.98315 \cdot W_{ij}^2 + 3.13160 \cdot W_{ij} + 0.33193 & \textrm{if } W_{ij} \leq 0.9 \\ \\
308097.45501 \cdot (W_{ij}-0.9)^4 - 42803.04696 \cdot (W_{ij}-0.9)^3 & \\
+ 2116.35304 \cdot (W_{ij}-0.9)^2 - 9.61869 \cdot (W_{ij}-0.9) + 2.86899 & \textrm{if } W_{ij} > 0.9.
\end{array} \right. 
\end{equation*}

In Group A, three home-away matches are played by Russia and another team. Then the expected number of goals scored by the host Russia (denoted by $R$) equals
\begin{equation*} \label{Exp_goals_home}
\lambda_{Rj}^{(h)} = 
\left\{ \begin{array}{ll}
-5.42301 \cdot W_{Rj}^4 + 15.49728 \cdot W_{Rj}^3 \\
- 12.6499 \cdot W_{Rj}^2 + 5.36198 \cdot W_{Rj} + 0.22863 & \textrm{if } W_{Rj} \leq 0.9 \\ \\
231098.16153 \cdot (W_{Rj}-0.9)^4 - 30953.10199 \cdot (W_{Rj}-0.9)^3 & \\
+ 1347.51495 \cdot (W_{Rj}-0.9)^2 - 1.63074 \cdot (W_{Rj}-0.9) + 2.54747 & \textrm{if } W_{Rj} > 0.9,
\end{array} \right.
\end{equation*}
and the expected number of goals scored by the away team $j$ is
\begin{equation*} \label{Exp_goals_away}
\lambda_{Rj}^{(a)} = 
\left\{ \begin{array}{ll}
90173.57949 \cdot (W_{Rj} - 0.1)^4 + 10064.38612 \cdot (W_{Rj} - 0.1)^3 \\
+ 218.6628 \cdot (W_{Rj} - 0.1)^2 - 11.06198 \cdot (W_{Rj} - 0.1) + 2.28291 & \textrm{if } W_{Rj} < 0.1 \\ \\
-1.25010 \cdot W_{Rj}^4 -  1.99984 \cdot W_{Rj}^3 & \\
+ 6.54946 \cdot W_{Rj}^2 - 5.83979 \cdot W_{Rj} + 2.80352 & \textrm{if } W_{Rj} \geq 0.1.
\end{array} \right.
\end{equation*}

The same simulation methodology has recently been applied in some studies on tournament design \citep{Csato2022a, Csato2023d, Stronka2024}. 
Note that even if this Elo-based approach is not necessarily the best available simulation model, it is only used to compare the outcomes under a uniform and a non-uniform draw. Thus, our assumptions seem to be adequate for comparative purposes as emphasised by \citet[p.~534]{Appleton1995} with the following words: ``\emph{Since our intention is to compare tournament designs, and not to estimate the chance of the best player winning any particular tournament, we may within reason take whatever model of determining winners that we please}''.

Last but not least, a ranking should be established based on the results of group matches. According to the rules of the 2018 FIFA World Cup, the first tie-breaking criteria are as follows \citep[Article~20.6]{FIFA2018f}:
(1) higher number of points obtained in all group matches;
(2) superior goal difference in all group matches;
(3) higher number of goals scored in all group matches.
In contrast to the official regulation, if some teams are still tied after considering (1)--(3), their ranking is decided by drawing of lots in our simulation.

Thus, a simulation run consists of the following steps for any draw mechanism (uniform or FIFA World Cup draw procedure with a given draw order):
\begin{enumerate}
\item
The eight groups are drawn according to the draw mechanism used (1 million times);
\item
The outcomes of all group matches are generated (10 times for each simulated draw);
\item
The group rankings are determined.
\end{enumerate}
The results for any team are the number of times it finishes on the first and the second place in its group, which are at most 10 million, respectively.

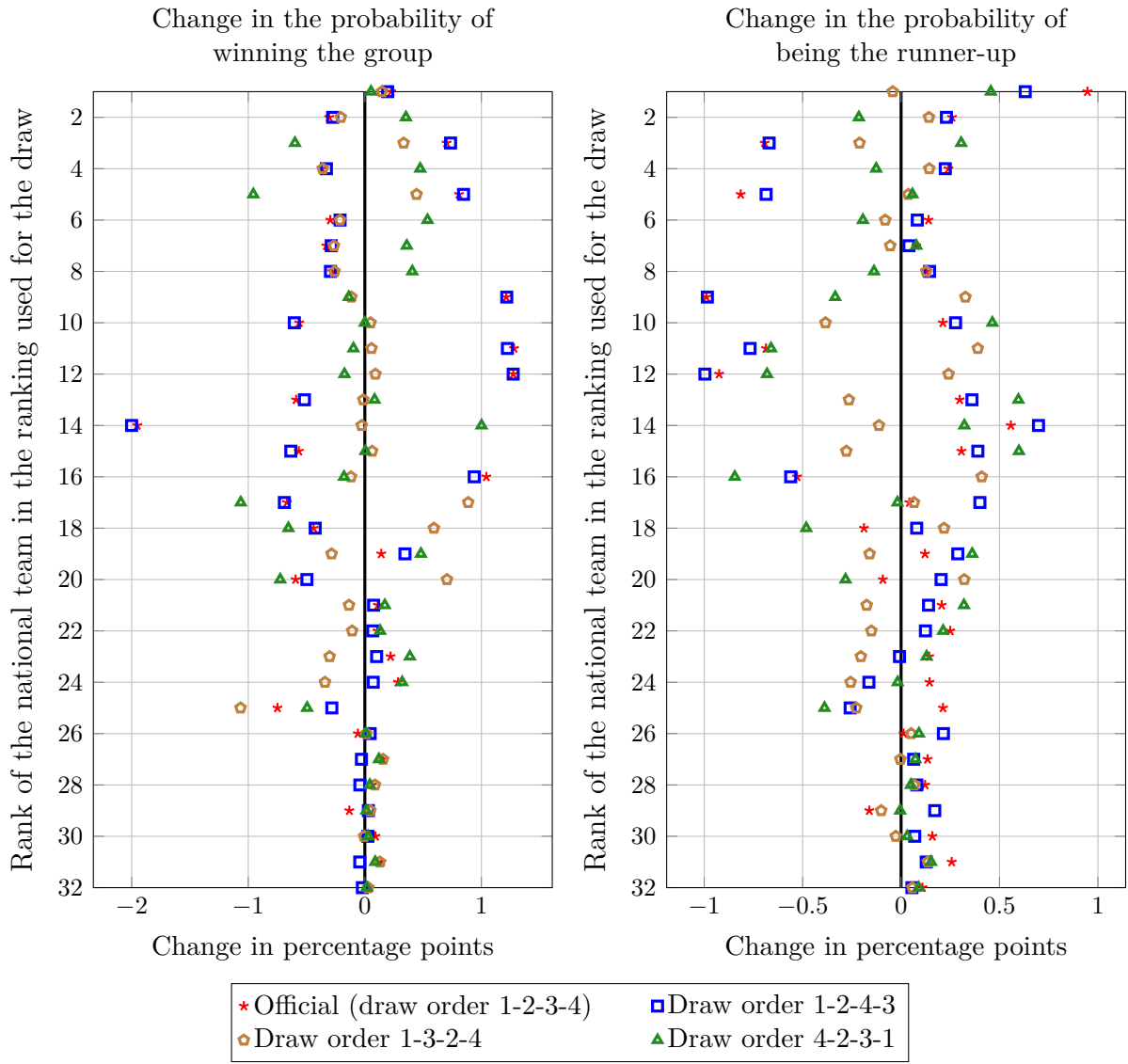
\begin{figure}[t!]
\centering

\begin{tikzpicture}
\begin{axis}[
name = axis1,
width = 0.5\textwidth, 
height = 0.8\textwidth,
title = {Change in the probability of \\ winning the group},
title style = {font = \small, align = center},
xmajorgrids,
ymajorgrids,
scaled x ticks = false,
xlabel = {Change in percentage points},
xlabel style = {align=center, font=\small},
xticklabel style = {/pgf/number format/fixed,/pgf/number format/precision=5},
extra x ticks = 0,
extra x tick labels = ,
extra x tick style = {grid = major, major grid style = {black,very thick}},
ymin = 1,
ymax = 32,
ylabel = {Rank of the national team in the ranking used for the draw},
ylabel style = {align=center, font=\small},
y dir = reverse,
legend style = {font=\small,at={(0.3,-0.12)},anchor=north west,legend columns=2},
legend entries = {Official (draw order 1-2-3-4)$\qquad$, Draw order 1-2-4-3, Draw order 1-3-2-4$\qquad \qquad \qquad$,Draw order 4-2-3-1},
]
\addplot [red, only marks, mark = star, thick] coordinates{
(0.22069,1)
(-0.33804,2)
(0.69158,3)
(-0.31077,4)
(0.75243,5)
(-0.32415,6)
(-0.25003,7)
(-0.32854,8)
(1.31955,9)
(-0.58276,10)
(1.20936,11)
(1.31114,12)
(-0.64842,13)
(-1.94395,14)
(-0.62014,15)
(1.07625,16)
(-0.60507,17)
(-0.43728,18)
(0.12756,19)
(-0.48676,20)
(0.10195,21)
(0.09515,22)
(0.1919,23)
(0.28444,24)
(-0.71897,25)
(-0.11627,26)
(0.14137,27)
(0.08633,28)
(-0.13983,29)
(0.09682,30)
(0.11184,31)
(0.03262,32)
};
\addplot [blue, only marks, mark = square, very thick] coordinates{
(0.05475,1)
(-0.3275,2)
(0.71369,3)
(-0.29306,4)
(0.80899,5)
(-0.29514,6)
(-0.24345,7)
(-0.33505,8)
(1.26955,9)
(-0.53484,10)
(1.14694,11)
(1.26764,12)
(-0.61623,13)
(-2.00817,14)
(-0.6246,15)
(1.01303,16)
(-0.54301,17)
(-0.38211,18)
(0.37311,19)
(-0.45392,20)
(0.06021,21)
(0.06418,22)
(0.11186,23)
(-0.00779,24)
(-0.30596,25)
(0.04235,26)
(-0.01019,27)
(-0.01493,28)
(0.06493,29)
(0.0088,30)
(0.00176,31)
(-0.00584,32)
};
\addplot [brown, only marks, mark = pentagon, very thick] coordinates{
(0.14553,1)
(-0.22974,2)
(0.26829,3)
(-0.22205,4)
(0.46747,5)
(-0.2348,6)
(-0.17574,7)
(-0.18954,8)
(-0.0738,9)
(0.02922,10)
(-0.06881,11)
(-0.02581,12)
(-0.05076,13)
(0.04104,14)
(-0.02116,15)
(-0.12732,16)
(1.00551,17)
(0.59523,18)
(-0.2379,19)
(0.71297,20)
(-0.14217,21)
(-0.12646,22)
(-0.30846,23)
(-0.37253,24)
(-1.15906,25)
(0.01415,26)
(0.14886,27)
(0.0939,28)
(0.04373,29)
(0.02866,30)
(0.13748,31)
(0.03407,32)
};
\addplot [ForestGreen, only marks, mark = triangle, very thick] coordinates{
(0.06946,1)
(0.38376,2)
(-0.70258,3)
(0.42832,4)
(-0.98181,5)
(0.4549,6)
(0.38627,7)
(0.46589,8)
(-0.13523,9)
(0.01519,10)
(-0.11257,11)
(-0.15188,12)
(-0.02955,13)
(1.07478,14)
(-0.04713,15)
(-0.10391,16)
(-1.00463,17)
(-0.60771,18)
(0.50539,19)
(-0.77291,20)
(0.17071,21)
(0.15879,22)
(0.32377,23)
(0.28647,24)
(-0.47375,25)
(0.00507,26)
(0.13824,27)
(0.07591,28)
(0.00465,29)
(0.04173,30)
(0.11431,31)
(0.02005,32)
};
\end{axis}

\begin{axis}[
at = {(axis1.south east)},
xshift = 0.1\textwidth,
width = 0.5\textwidth, 
height = 0.8\textwidth,
title = {Change in the probability of \\ being the runner-up},
title style = {font = \small, align = center},
xmajorgrids,
ymajorgrids,
scaled x ticks = false,
xlabel = {Change in percentage points},
xlabel style = {align=center, font=\small},
xticklabel style = {/pgf/number format/fixed,/pgf/number format/precision=5},
extra x ticks = 0,
extra x tick labels = ,
extra x tick style = {grid = major, major grid style = {black,very thick}},
ymin = 1,
ymax = 32,
ylabel = {Rank of the national team in the ranking used for the draw},
ylabel style = {align=center, font=\small},
y dir = reverse,
]
\addplot [red, only marks, mark = star, thick] coordinates{
(0.94566,1)
(0.25884,2)
(-0.67493,3)
(0.1984,4)
(-0.79428,5)
(0.17518,6)
(0.07081,7)
(0.21046,8)
(-1.09049,9)
(0.30969,10)
(-0.72354,11)
(-0.93452,12)
(0.32038,13)
(0.52811,14)
(0.27676,15)
(-0.57425,16)
(0.01768,17)
(-0.21563,18)
(0.12308,19)
(-0.12249,20)
(0.25168,21)
(0.21624,22)
(0.25391,23)
(0.32892,24)
(0.24912,25)
(-0.12032,26)
(0.09375,27)
(0.13793,28)
(-0.11061,29)
(0.15112,30)
(0.13603,31)
(0.10731,32)
};
\addplot [blue, only marks, mark = square, very thick] coordinates{
(0.7264,1)
(0.23828,2)
(-0.65391,3)
(0.18955,4)
(-0.74304,5)
(0.15412,6)
(0.05811,7)
(0.20883,8)
(-1.06551,9)
(0.34719,10)
(-0.74208,11)
(-0.93663,12)
(0.36128,13)
(0.64976,14)
(0.34246,15)
(-0.60109,16)
(0.28576,17)
(0.08152,18)
(0.30776,19)
(0.16483,20)
(0.08464,21)
(0.07736,22)
(0.04815,23)
(0.01392,24)
(-0.15956,25)
(0.17715,26)
(0.01735,27)
(0.04247,28)
(0.19995,29)
(0.0487,30)
(0.04732,31)
(0.02896,32)
};
\addplot [brown, only marks, mark = pentagon, very thick] coordinates{
(-0.11329,1)
(0.14327,2)
(-0.13743,3)
(0.05906,4)
(0.01729,5)
(0.0253,6)
(-0.08581,7)
(0.03673,8)
(0.23575,9)
(-0.29082,10)
(0.35897,11)
(0.27928,12)
(-0.25236,13)
(-0.18517,14)
(-0.27356,15)
(0.38751,16)
(-0.00549,17)
(0.24316,18)
(-0.19247,19)
(0.2024,20)
(-0.18704,21)
(-0.17064,22)
(-0.17021,23)
(-0.01986,24)
(-0.13317,25)
(-0.0066,26)
(0.04485,27)
(0.08634,28)
(-0.01432,29)
(0.00067,30)
(0.06293,31)
(0.05473,32)
};
\addplot [ForestGreen, only marks, mark = triangle, very thick] coordinates{
(0.42856,1)
(-0.22789,2)
(0.3771,3)
(-0.12149,4)
(0.01899,5)
(-0.09223,6)
(0.08087,7)
(-0.17679,8)
(-0.33816,9)
(0.56553,10)
(-0.80048,11)
(-0.62141,12)
(0.55691,13)
(0.27076,14)
(0.63126,15)
(-0.8944,16)
(-0.15356,17)
(-0.35751,18)
(0.32359,19)
(-0.32413,20)
(0.22879,21)
(0.2247,22)
(0.24541,23)
(0.14004,24)
(-0.42285,25)
(0.01306,26)
(0.05324,27)
(0.11608,28)
(0.01437,29)
(0.03085,30)
(0.12685,31)
(0.08394,32)
};
\end{axis}
\end{tikzpicture}

\caption{The effect of different draw procedures on being the \\ group winner and the runner-up in the 2018 FIFA World Cup}
\label{Fig4}

\end{figure}


Figure~\ref{Fig4} shows how the FIFA World Cup draw mechanism distorts the chances of winning the group and being the runner-up. The official rule (draw order 1-2-3-4) increases the probability of winning the group by more than 1 percentage point for the four UEFA teams in Pot 2 (Spain [9], Switzerland [11], England [12], Croatia [16]) mostly at the expense of Mexico (14) and Serbia (25). Fortunately, these effects are somewhat mitigated by taking the likelihood of obtaining the second position into account. While the draw order 1-2-4-3 does not differ much from the traditional order of 1-2-3-4, the draw order 1-3-2-4 strongly reduces the distortions in the case of the above countries except for Serbia. Among the four draw orders, 4-2-3-1 is the less favourable for the weakest European team in Pot 2 (Croatia [16]) and the three UEFA members in Pot 3 (Denmark [17], Iceland [18], Sweden [20]).

\begin{figure}[t!]
\centering

\begin{tikzpicture}
\begin{axis}[
name = axis1,
width = 0.8\textwidth, 
height = 0.8\textwidth,
title = {Absolute changes in the probability of qualification},
title style = {font = \small},
xmajorgrids,
ymajorgrids,
scaled x ticks = false,
xlabel = {Change in percentage points},
xlabel style = {align=center, font=\small},
xticklabel style = {/pgf/number format/fixed,/pgf/number format/precision=5},
extra x ticks = 0,
extra x tick labels = ,
extra x tick style = {grid = major, major grid style = {black,very thick}},
symbolic y coords = {Russia (1),Germany (2),Brazil (3),Portugal (4),Argentina (5),Belgium (6),Poland (7),France (8),Spain (9),Peru (10),Switzerland (11),England (12),Colombia (13),Mexico (14),Uruguay (15),Croatia (16),Denmark (17),Iceland (18),Costa Rica (19),Sweden (20),Tunisia (21),Egypt (22),Senegal (23),Iran (24),Serbia (25),Nigeria (26),Australia (27),Japan (28),Morocco (29),Panama (30),South Korea (31),Saudi Arabia (32)},
ytick = data,
y dir = reverse,
enlarge y limits = 0.02,
legend style = {font=\small,at={(0,-0.12)},anchor=north west,legend columns=2},
legend entries = {Official (draw order 1-2-3-4)$\qquad$, Draw order 1-2-4-3, Draw order 1-3-2-4$\qquad \qquad \qquad$,Draw order 4-2-3-1},
]
\addplot [red, only marks, mark = star, thick] coordinates{
(1.16635,{Russia (1)})
(-0.0792,{Germany (2)})
(0.01665,{Brazil (3)})
(-0.11237,{Portugal (4)})
(-0.04185,{Argentina (5)})
(-0.14897,{Belgium (6)})
(-0.17922,{Poland (7)})
(-0.11808,{France (8)})
(0.22906,{Spain (9)})
(-0.27307,{Peru (10)})
(0.48582,{Switzerland (11)})
(0.37662,{England (12)})
(-0.32804,{Colombia (13)})
(-1.41584,{Mexico (14)})
(-0.34338,{Uruguay (15)})
(0.502,{Croatia (16)})
(-0.58739,{Denmark (17)})
(-0.65291,{Iceland (18)})
(0.25064,{Costa Rica (19)})
(-0.60925,{Sweden (20)})
(0.35363,{Tunisia (21)})
(0.31139,{Egypt (22)})
(0.44581,{Senegal (23)})
(0.61336,{Iran (24)})
(-0.46985,{Serbia (25)})
(-0.23659,{Nigeria (26)})
(0.23512,{Australia (27)})
(0.22426,{Japan (28)})
(-0.25044,{Morocco (29)})
(0.24794,{Panama (30)})
(0.24787,{South Korea (31)})
(0.13993,{Saudi Arabia (32)})
};
\addplot [blue, only marks, mark = square, very thick] coordinates{
(0.78115,{Russia (1)})
(-0.08922,{Germany (2)})
(0.05978,{Brazil (3)})
(-0.10351,{Portugal (4)})
(0.06595,{Argentina (5)})
(-0.14102,{Belgium (6)})
(-0.18534,{Poland (7)})
(-0.12622,{France (8)})
(0.20404,{Spain (9)})
(-0.18765,{Peru (10)})
(0.40486,{Switzerland (11)})
(0.33101,{England (12)})
(-0.25495,{Colombia (13)})
(-1.35841,{Mexico (14)})
(-0.28214,{Uruguay (15)})
(0.41194,{Croatia (16)})
(-0.25725,{Denmark (17)})
(-0.30059,{Iceland (18)})
(0.68087,{Costa Rica (19)})
(-0.28909,{Sweden (20)})
(0.14485,{Tunisia (21)})
(0.14154,{Egypt (22)})
(0.16001,{Senegal (23)})
(0.00613,{Iran (24)})
(-0.46552,{Serbia (25)})
(0.2195,{Nigeria (26)})
(0.00716,{Australia (27)})
(0.02754,{Japan (28)})
(0.26488,{Morocco (29)})
(0.0575,{Panama (30)})
(0.04908,{South Korea (31)})
(0.02312,{Saudi Arabia (32)})
};
\addplot [brown, only marks, mark = pentagon, very thick] coordinates{
(0.03224,{Russia (1)})
(-0.08647,{Germany (2)})
(0.13086,{Brazil (3)})
(-0.16299,{Portugal (4)})
(0.48476,{Argentina (5)})
(-0.2095,{Belgium (6)})
(-0.26155,{Poland (7)})
(-0.15281,{France (8)})
(0.16195,{Spain (9)})
(-0.2616,{Peru (10)})
(0.29016,{Switzerland (11)})
(0.25347,{England (12)})
(-0.30312,{Colombia (13)})
(-0.14413,{Mexico (14)})
(-0.29472,{Uruguay (15)})
(0.26019,{Croatia (16)})
(1.00002,{Denmark (17)})
(0.83839,{Iceland (18)})
(-0.43037,{Costa Rica (19)})
(0.91537,{Sweden (20)})
(-0.32921,{Tunisia (21)})
(-0.2971,{Egypt (22)})
(-0.47867,{Senegal (23)})
(-0.39239,{Iran (24)})
(-1.29223,{Serbia (25)})
(0.00755,{Nigeria (26)})
(0.19371,{Australia (27)})
(0.18024,{Japan (28)})
(0.02941,{Morocco (29)})
(0.02933,{Panama (30)})
(0.20041,{South Korea (31)})
(0.0888,{Saudi Arabia (32)})
};
\addplot [ForestGreen, only marks, mark = triangle, very thick] coordinates{
(0.49802,{Russia (1)})
(0.15587,{Germany (2)})
(-0.32548,{Brazil (3)})
(0.30683,{Portugal (4)})
(-0.96282,{Argentina (5)})
(0.36267,{Belgium (6)})
(0.46714,{Poland (7)})
(0.2891,{France (8)})
(-0.47339,{Spain (9)})
(0.58072,{Peru (10)})
(-0.91305,{Switzerland (11)})
(-0.77329,{England (12)})
(0.52736,{Colombia (13)})
(1.34554,{Mexico (14)})
(0.58413,{Uruguay (15)})
(-0.99831,{Croatia (16)})
(-1.15819,{Denmark (17)})
(-0.96522,{Iceland (18)})
(0.82898,{Costa Rica (19)})
(-1.09704,{Sweden (20)})
(0.3995,{Tunisia (21)})
(0.38349,{Egypt (22)})
(0.56918,{Senegal (23)})
(0.42651,{Iran (24)})
(-0.8966,{Serbia (25)})
(0.01813,{Nigeria (26)})
(0.19148,{Australia (27)})
(0.19199,{Japan (28)})
(0.01902,{Morocco (29)})
(0.07258,{Panama (30)})
(0.24116,{South Korea (31)})
(0.10399,{Saudi Arabia (32)})
};
\end{axis}
\end{tikzpicture}

\caption{The absolute effect of different draw procedures on the probability \\ of qualification for the knockout stage in the 2018 FIFA World Cup}
\label{Fig5}

\end{figure}
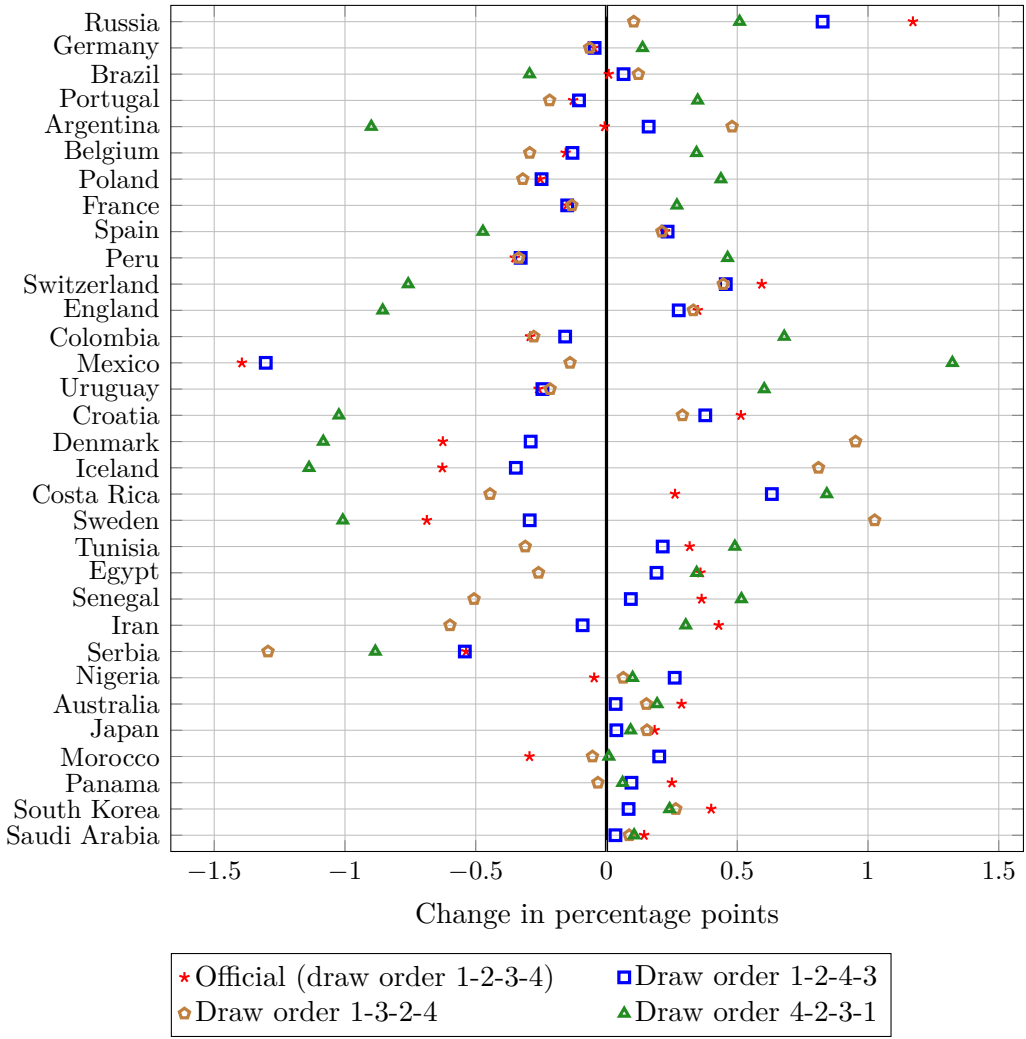


\begin{table}[t!]
  \centering
  \caption{Summary statistics on the absolute changes of qualifying \\ probabilities by four draw procedures in the 2018 FIFA World Cup}
  \label{Table4}
    \rowcolors{1}{gray!20}{}
\centerline{
\begin{threeparttable}
    \begin{tabularx}{1.1\textwidth}{l CCC CCC CCC CCC} \toprule \hiderowcolors
    Draw order & \multicolumn{3}{c}{\textbf{1-2-3-4}} & \multicolumn{3}{c}{\textbf{1-2-4-3}} & \multicolumn{3}{c}{\textbf{1-3-2-4}} & \multicolumn{3}{c}{\textbf{4-2-3-1}} \\ \midrule
    Change (pp) & GW    & RU    & Q     & GW    & RU    & Q     & GW    & RU    & Q     & GW    & RU    & Q \\ \bottomrule \showrowcolors
    $> 1$ & 4     & 0     & 1     & 4     & 0     & 0     & 1     & 0     & 1     & 1     & 0     & 1 \\
    $0.5$ -- $1$ & 2     & 2     & 2     & 2     & 2     & 2     & 2     & 0     & 2     & 1     & 3     & 5 \\
    $0.25$ -- $0.5$ & 1     & 7     & 6     & 1     & 5     & 4     & 2     & 3     & 4     & 7     & 4     & 8 \\
    $-0.25$ -- $0.25$ & 12    & 17    & 14    & 13    & 19    & 19    & 24    & 26    & 15    & 17    & 18    & 8 \\
    $-0.5$ -- $-0.25$ & 7     & 0     & 5     & 7     & 0     & 6     & 2     & 3     & 9     & 1     & 4     & 2 \\
    $-1$ -- $-0.5$ & 5     & 5     & 3     & 4     & 5     & 0     & 0     & 0     & 0     & 4     & 3     & 6 \\
    $< -1$ & 1     & 1     & 1     & 1     & 1     & 1     & 1     & 0     & 1     & 1     & 0     & 2 \\ \toprule
    \end{tabularx}
\begin{tablenotes} \footnotesize
\item
Column GW/RU/Q shows the number of teams for which the absolute change in percentage points of the probability of winning the group/being the runner-up/qualification by the given draw order (the probability according to the official draw mechanism with the given draw order minus the probability according to a uniform draw) is within the given interval.
\end{tablenotes}
\end{threeparttable}
}
\end{table}

The probabilities of being the group winner or the runner-up are aggregated in Figure~\ref{Fig5}. The draw order 1-2-4-3 is somewhat better than the official 1-2-3-4: despite the increased impact on Costa Rica, it is less unfair for Russia, Croatia, Denmark, Iceland, and Sweden. The draw order 1-3-2-4 can be chosen if the bias for Mexico should be reduced, however, that is achieved at the expense of Serbia. The draw order 4-2-3-1 is not worth implementing because of the high distortions for several countries. 
These observations are reinforced by the summary statistics provided in Table~\ref{Table4}.

\begin{figure}[t!]
\centering

\begin{tikzpicture}
\begin{axis}[
name = axis1,
width = 0.8\textwidth, 
height = 0.8\textwidth,
title = {Relative changes in the probability of qualification},
title style = {font = \small},
xmajorgrids,
ymajorgrids,
scaled x ticks = false,
xlabel = {Change in \%},
xlabel style = {align=center, font=\small},
xticklabel style = {/pgf/number format/fixed,/pgf/number format/precision=5},
extra x ticks = 0,
extra x tick labels = ,
extra x tick style = {grid = major, major grid style = {black,very thick}},
symbolic y coords = {Russia (1),Germany (2),Brazil (3),Portugal (4),Argentina (5),Belgium (6),Poland (7),France (8),Spain (9),Peru (10),Switzerland (11),England (12),Colombia (13),Mexico (14),Uruguay (15),Croatia (16),Denmark (17),Iceland (18),Costa Rica (19),Sweden (20),Tunisia (21),Egypt (22),Senegal (23),Iran (24),Serbia (25),Nigeria (26),Australia (27),Japan (28),Morocco (29),Panama (30),South Korea (31),Saudi Arabia (32)},
ytick = data,
y dir = reverse,
enlarge y limits = 0.02,
legend style = {font=\small,at={(0,-0.12)},anchor=north west,legend columns=2},
legend entries = {Official (draw order 1-2-3-4)$\qquad$, Draw order 1-2-4-3, Draw order 1-3-2-4$\qquad \qquad \qquad$,Draw order 4-2-3-1},
]
\addplot [red, only marks, mark = star, thick] coordinates{
(2.08791531407611,{Russia (1)})
(-0.0845035438140185,{Germany (2)})
(0.0172588275793473,{Brazil (3)})
(-0.135113865626091,{Portugal (4)})
(-0.050109864530945,{Argentina (5)})
(-0.189309075881461,{Belgium (6)})
(-0.298181164773292,{Poland (7)})
(-0.138426704537065,{France (8)})
(0.254808982812027,{Spain (9)})
(-0.373637428161644,{Peru (10)})
(0.704626046507184,{Switzerland (11)})
(0.478737359767645,{England (12)})
(-0.437604944012737,{Colombia (13)})
(-2.40932978858635,{Mexico (14)})
(-0.496916377589884,{Uruguay (15)})
(0.80531507952486,{Croatia (16)})
(-1.22129497312335,{Denmark (17)})
(-2.10917196320555,{Iceland (18)})
(0.943263796606675,{Costa Rica (19)})
(-1.66896454617999,{Sweden (20)})
(2.37647862089914,{Tunisia (21)})
(2.30317536277655,{Egypt (22)})
(1.57302696809205,{Senegal (23)})
(1.76379177272921,{Iran (24)})
(-1.47176288908449,{Serbia (25)})
(-1.53733490495542,{Nigeria (26)})
(0.924262883447047,{Australia (27)})
(1.30933336447179,{Japan (28)})
(-1.13184216758351,{Morocco (29)})
(1.86537743911981,{Panama (30)})
(1.17578472613566,{South Korea (31)})
(1.89713727909326,{Saudi Arabia (32)})
};
\addplot [blue, only marks, mark = square, very thick] coordinates{
(1.39835816657996,{Russia (1)})
(-0.0951945224632156,{Germany (2)})
(0.0619659286902552,{Brazil (3)})
(-0.124460587620867,{Portugal (4)})
(0.0789664412381264,{Argentina (5)})
(-0.179206322620684,{Belgium (6)})
(-0.308363447601179,{Poland (7)})
(-0.147969331357278,{France (8)})
(0.226976446577165,{Spain (9)})
(-0.256758572507165,{Peru (10)})
(0.58720287593943,{Switzerland (11)})
(0.420760590135116,{England (12)})
(-0.340102976698109,{Colombia (13)})
(-2.31160136605377,{Mexico (14)})
(-0.408293979769381,{Uruguay (15)})
(0.660839629202137,{Croatia (16)})
(-0.534871434372364,{Denmark (17)})
(-0.971031230062269,{Iceland (18)})
(2.56240033991217,{Costa Rica (19)})
(-0.791926074116001,{Sweden (20)})
(0.973426825318113,{Tunisia (21)})
(1.0468911681409,{Egypt (22)})
(0.564590397623221,{Senegal (23)})
(0.0176275654865554,{Iran (24)})
(-1.45819955331832,{Serbia (25)})
(1.42628602915471,{Nigeria (26)})
(0.028146147692576,{Australia (27)})
(0.160791228295509,{Japan (28)})
(1.19710251297525,{Morocco (29)})
(0.432601446920189,{Panama (30)})
(0.232813629558803,{South Korea (31)})
(0.313455398360873,{Saudi Arabia (32)})
};
\addplot [brown, only marks, mark = pentagon, very thick] coordinates{
(0.0577137134872174,{Russia (1)})
(-0.0922603716363435,{Germany (2)})
(0.135645055677602,{Brazil (3)})
(-0.195979433642401,{Portugal (4)})
(0.580436270729279,{Argentina (5)})
(-0.266229787186456,{Belgium (6)})
(-0.435159489155534,{Poland (7)})
(-0.179141130761418,{France (8)})
(0.180155045692865,{Spain (9)})
(-0.357943205797362,{Peru (10)})
(0.420843715068386,{Switzerland (11)})
(0.32219626833494,{England (12)})
(-0.404361695613775,{Colombia (13)})
(-0.245265497816805,{Mexico (14)})
(-0.426498907342565,{Uruguay (15)})
(0.417400260042977,{Croatia (16)})
(2.07923083304589,{Denmark (17)})
(2.70834982192323,{Iceland (18)})
(-1.61966342222157,{Costa Rica (19)})
(2.50754218569844,{Sweden (20)})
(-2.21237034976163,{Tunisia (21)})
(-2.19748033103477,{Egypt (22)})
(-1.68897247440978,{Senegal (23)})
(-1.12836548470917,{Iran (24)})
(-4.04779431342269,{Serbia (25)})
(0.0490590410939351,{Nigeria (26)})
(0.7614790879233,{Australia (27)})
(1.0523242915027,{Japan (28)})
(0.132915980468895,{Morocco (29)})
(0.220664355446409,{Panama (30)})
(0.95065565403174,{South Korea (31)})
(1.20392903868707,{Saudi Arabia (32)})
};
\addplot [ForestGreen, only marks, mark = triangle, very thick] coordinates{
(0.891519342149594,{Russia (1)})
(0.166307668867316,{Germany (2)})
(-0.337381573604956,{Brazil (3)})
(0.368932877013917,{Portugal (4)})
(-1.15285017366031,{Argentina (5)})
(0.460876166677382,{Belgium (6)})
(0.777214313760721,{Poland (7)})
(0.338915652791871,{France (8)})
(-0.526604489537152,{Spain (9)})
(0.794590131768524,{Peru (10)})
(-1.32427403516401,{Switzerland (11)})
(-0.982961109167679,{England (12)})
(0.703497571255207,{Colombia (13)})
(2.28970053377111,{Mexico (14)})
(0.84531354080486,{Uruguay (15)})
(-1.60150218533958,{Croatia (16)})
(-2.40809619660148,{Denmark (17)})
(-3.11806368768323,{Iceland (18)})
(3.11980059891079,{Costa Rica (19)})
(-3.00520453958356,{Sweden (20)})
(2.6847360491169,{Tunisia (21)})
(2.83645820312528,{Egypt (22)})
(2.0083342448546,{Senegal (23)})
(1.22648172196873,{Iran (24)})
(-2.80851890252879,{Serbia (25)})
(0.117806677487819,{Nigeria (26)})
(0.752712899465968,{Australia (27)})
(1.12092621352422,{Japan (28)})
(0.0859592638054574,{Morocco (29)})
(0.546055878564644,{Panama (30)})
(1.14395547889972,{South Korea (31)})
(1.40987140465167,{Saudi Arabia (32)})
};
\end{axis}
\end{tikzpicture}

\caption{The relative effect of different draw procedures on the probability \\ of qualification for the knockout stage in the 2018 FIFA World Cup}
\label{Fig6}

\end{figure}
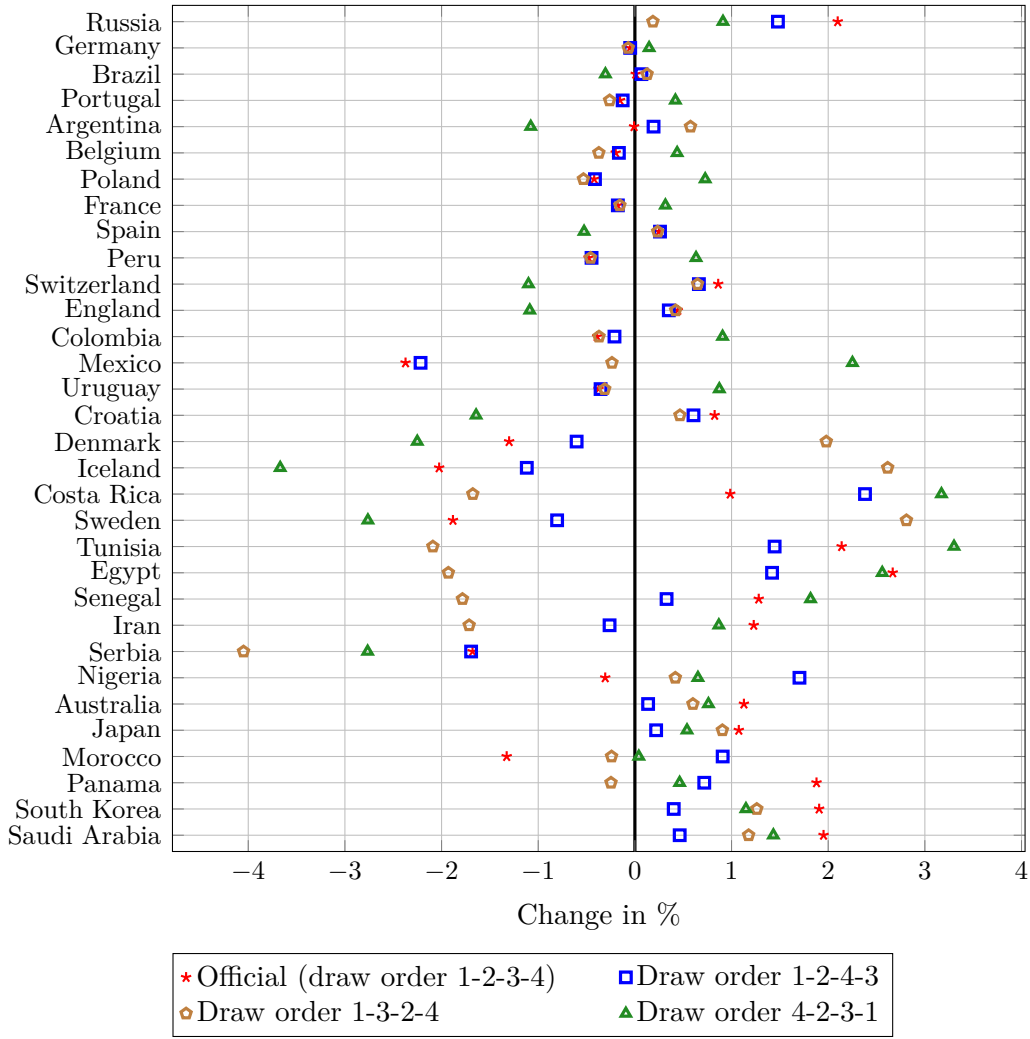


\begin{table}[t!]
  \centering
  \caption{Summary statistics on the relative changes of qualifying \\ probabilities by four draw procedures in the 2018 FIFA World Cup}
  \label{Table5}
    \rowcolors{1}{}{gray!20}
\begin{threeparttable}
    \begin{tabularx}{0.8\textwidth}{l CCCC} \toprule
    Change & 1-2-3-4 & 1-2-4-3 & 1-3-2-4 & 4-2-3-1 \\ \bottomrule
    $> 3\%$ & 0     & 0     & 0     & 1 \\
    $2\%$ -- $3\%$ & 3     & 1     & 3     & 4 \\
    $1\%$ -- $2\%$ & 6     & 4     & 2     & 4 \\
    $-1\%$ -- $1\%$ & 16    & 25    & 21    & 16 \\
    $-2\%$ -- $-1\%$ & 5     & 1     & 3     & 3 \\
    $-3\%$ -- $-2\%$ & 2     & 1     & 2     & 2 \\
    $< -3\%$ & 0     & 0     & 1     & 2 \\ \toprule
    \end{tabularx}
\begin{tablenotes} \footnotesize
\item
The cells show the number of national teams for which the relative change in percentage of the probability of qualification by the given draw order (one minus (the probability according to the official draw mechanism with the given draw order divided by the probability according to a uniform draw)) is within the appropriate interval.
\end{tablenotes}
\end{threeparttable}
\end{table}

Finally, Figure~\ref{Fig6} uncovers the relative changes in the probability of qualification for the knockout stage as the same absolute distortions can be more costly for the weaker teams. In this respect, the draw order 1-2-4-3 can be a reasonable alternative to the official 1-2-3-4 as it is less biased for almost all teams except for Costa Rica as shown by Table~\ref{Table5}. The other two draw orders are especially unfavourable for Serbia and the three UEFA (draw order 4-2-3-1) or the three CAF (draw order 1-3-2-4) teams in Pot 3.

Based on the arguments above, the official draw order of Pot 1, Pot 2, Pot 3, Pot 4 has been a lucky choice for the 2018 FIFA World Cup draw.
But a change in the continental allocation of the teams playing in the FIFA World Cup may lead to a highly unfair draw in the future if the current draw procedure is applied.

\section{Discussion} \label{Sec5}

Our paper has analysed the unfairness of the 2018 FIFA World Cup draw.
First, the connection of the standard constrained group draw procedure---used by the FIBA, FIFA, and UEFA---to permutation generation has been presented.
Second, we have examined how this procedure departs from a random draw among all feasible allocations and considered all alternatives by relabelling the pots. The official draw order (Pot 1, Pot 2, Pot 3, Pot 4) has turned out to be an optimal choice among the 24 alternative rules with respect to natural measures of fairness.

Even though constrained group draw is currently used only in basketball and football, draw constraints offer the unique solution to maximise the number of intercontinental games in a group stage. This can be advantageous in several tournaments; for instance, the 2023 World Men's Handball Championship contained eight groups, one with two of the three African teams, and another with two of the four South American teams \citep{DevriesereCsatoGoossens2024}. Any governing body will likely adopt the procedure of FIFA if they introduce draw constraints.

We think there is at least some scope to modify the current draw order because the draw of certain sports tournaments has not been started with the strongest teams:
\begin{itemize}
\item
The lowest-ranked teams were drawn first in the 2020/21 \citep{UEFA2020d} and 2022/23 \citep{UEFA2021i} UEFA Nations League, as well as in the 2019 World Men's Handball Championship \citep{IHF2018b};
\item
The runners-up were drawn first in the UEFA Champions League Round of 16 \citep{BoczonWilson2023, KlossnerBecker2013}, and the unseeded teams were drawn first in the UEFA Europa League Round of 32 \citep{Csato2022d};
\item
The unusual draw order of Pot 4, Pot 3, Pot 1, Pot 2 was followed in the draw of the 2021 World Men's Handball Championship \citep{IHF2020}.
\end{itemize}

According to our findings, the non-uniform draw distorts the probability of qualification for eight (two) countries by more than 0.5 (1) percentage points (see Table~\ref{Table4}). Although the bias has not exceeded 1.5 percentage points for any national team in the case of the 2018 FIFA World Cup, it has exceeded 2\% for Russia, Mexico, Iceland, Tunisia, and Egypt in relative terms. 
To conclude, it is the responsibility of policymakers to decide whether these values justify the consideration of fairer draw mechanisms such as the algorithms proposed by \citet{Guyon2014a} and \citet{RobertsRosenthal2024}.


\section*{Acknowledgements}
\addcontentsline{toc}{section}{Acknowledgements}
\noindent
This paper could not have been written without \emph{my father} (also called \emph{L\'aszl\'o Csat\'o}), who has primarily coded the simulations in Python. \\
We are grateful to \emph{Julien Guyon} for inspiration, and to \emph{Julien Guyon}, \emph{Jeffrey S.~Rosenthal}, and \emph{Lajos R\'onyai} for useful advice. \\
The comments of the reviewers \emph{Martin Becker}, \emph{Florian Felice}, and \emph{Michael A.~Lapr\'e} have greatly improved the presentation of the results. \\
Eleven colleagues and anonymous reviewers provided valuable remarks and suggestions on earlier drafts. \\
We are indebted to the \href{https://en.wikipedia.org/wiki/Wikipedia_community}{Wikipedia community} for summarising important details of the sports competition discussed in the paper. \\
The research was supported by the National Research, Development and Innovation Office under Grant FK 145838, and by the J\'anos Bolyai Research Scholarship of the Hungarian Academy of Sciences.

\bibliographystyle{apalike}
\bibliography{All_references}

\end{document}